\documentclass[12pt]{article}

\usepackage[left=1in,right=1in,top=1in,bottom=1in]{geometry}
\usepackage{graphicx}
\usepackage{makecell} 
\usepackage{natbib}
\defcitealias{cea_2015}{DOT, CEA, and DOL, 2015}
\defcitealias{cea_2024}{CEA, 2024}
\defcitealias{bea_2003}{BEA, 2015}
\usepackage{setspace}
\usepackage{soul}               
\usepackage{caption}
\usepackage{subcaption}
\usepackage{booktabs}
\usepackage{threeparttable}
\usepackage{siunitx}

\usepackage{scalerel}
\usepackage{nicefrac}

\usepackage[]{appendix}

\usepackage[shortlabels,inline]{enumitem}
\setlist[description]{leftmargin=0em,labelindent=\parindent}

\usepackage{amsmath,amsthm}

\theoremstyle{definition}

\newtheorem{result}{Result}

\theoremstyle{remark}

\usepackage{mathtools}

\usepackage[T1]{fontenc}

\usepackage[scaled=.85]{beramono}

\usepackage{amsmath,amsthm}
\usepackage[libertine]{newtxmath}

\usepackage[scr=rsfso]{mathalfa}
\usepackage{bm}

\usepackage[semibold]{libertine}

\usepackage{josefin}

\usepackage{titlesec}

\titleformat{\section}[block]{\bf\Large}{\thesection\quad}{0pt}{}
\titleformat{\subsection}[block]{\bf\large}{\thesubsection\quad}{0pt}{}

\newenvironment{tablenotesminipage}[1][Note]{\begin{minipage}[t]{\linewidth}\footnotesize{\itshape#1: }}{\end{minipage}}
\newenvironment{figurenotes}[1][Note]{\begin{minipage}[t]{\linewidth}\footnotesize{\itshape#1: }}{\end{minipage}}

\newcommand{\tp}{{\cramped{^{\mathsf{{T}}}}}} 

\newcommand{\lsp}{\eta}

\newcommand{\hhincome}{\omega}

\usepackage[dvipsnames]{xcolor}
\definecolor{csub-blue}{RGB}{0, 53, 148}
\definecolor{csub-blue-hl}{RGB}{0, 53, 148}
\usepackage{tikz} 
\usepackage[unicode,hyperfootnotes=true,linktocpage,pagebackref]{hyperref}
 \hypersetup{%
   colorlinks = true,
   pdfborder = {0 0 0},
   citecolor = csub-blue,
   linkcolor = csub-blue,
   urlcolor  = csub-blue,
 }%
\makeatletter
\def\@fnsymbol#1{\ensuremath{\ifcase#1\or *\or \mathsection\or \mathparagraph\or \|\or **\or \dagger\dagger \or \ddagger\ddagger \else\@ctrerr\fi}}
\makeatother

\title{\textsf{\Large Employment, Input--Output Linkages, and the\\Energy Transition in California's Top Oil-Producing Region}\thanks{We gratefully acknowledge support from the Resilient Energy Economies Initiative.}}
\author{Rich Ryan\thanks{Corresponding author. Email: \href{mailto:richryan@csub.edu}{richryan@csub.edu}. Department of Economics, California State University, Bakersfield.
  ORCID: \href{https://orcid.org/0000-0002-6648-5413}{\nolinkurl{https://orcid.org/0000-0002-6648-5413}}.}
  \and
  Nyakundi Michieka\thanks{Email: \href{mailto:nmichieka@csub.edu}{nmichieka@csub.edu}.
    Department of Economics, California State University, Bakersfield.
  ORCID: \href{https://orcid.org/0000-0002-8725-2774}{\nolinkurl{https://orcid.org/0000-0002-8725-2774}}.}} 

\date{April 10, 2026}

\begin{document}
\maketitle

\begin{abstract}
  The US economy is transitioning away from fossil fuels toward sources of green energy.
  California policymakers have adopted the goal of carbon neutrality by 2045 or earlier.
  Within California, Kern County accounts for over 70 percent of oil produced within the state.
  To understand how the transition may affect opportunities in Kern, 
  we propose a structural vector autoregressive model that jointly explains the global crude-oil market and the evolution of employment within Kern.
  We use monthly data from the Quarterly Census of Employment and Wages. 
  While industries directly involved in the extraction of fossil fuels employ less than 2 percent of workers,
  the oil market is responsible for 11 percent of the variation in employment growth.
  Employment would be $6.4$ percent lower currently absent the influence of the global oil market. 
  We explain these large effects using a theoretical framework of production that relies on
  a network of input--output linkages.
  The findings may be useful to policymakers designing place-based policy aimed at helping vulnerable oil-dependent regions.
\end{abstract}

\vspace{3em}


\vspace{1em}


\vfill

\clearpage
\pagebreak

\section{Introduction}

Concern about the rise in Earth's temperature has led to action.
Between 2005 and 2021
US greenhouse gas emissions fell by 17 percent while
economic activity increased enough to allow real income per person to increase by 19 percent.
Yet, despite these gains,
the transition away from fossil fuels to sources of clean energy has not been
rapid enough to achieve the goal of the Paris Agreement---limiting
global warming to 1.5 degrees Celsius \citepalias[][212--213]{cea_2024}.
There is good reason to believe that
the pace of the transition away from fossil fuels will only accelerate
given the increased frequency of
wildfires, droughts, floods, hurricanes and other damaging climate events \citep{fernandez-villaverde_gillingham_scheidegger_2025}.
It is unknown what 
an accelerated transition would mean for work in oil-extracting regions.

In short,
here is the main issue:
The green energy transition is underway.
As the United States transitions away from fossil fuels,
regions across the United States like Kern County, California,
which accounts for $70$ percent of the oil produced within the state,
are concerned about lost tax revenue and lost jobs.
To what extent does Kern County's economy depend on oil?

We focus on Kern County, California, for several reasons.
In addition to US goals,
California policymakers have adopted the goal of carbon neutrality by 2045 or earlier
\citep{carb_2022}.
The state has recently flirted with ending new drilling of oil wells.\footnote{Sangmin Kim,
  ``Third time's the charm:
  Kern County seeks to allow oil drilling after courts overturn decision twice,''
\textit{KGET.com}, \href{https://www.kget.com/news/local-news/third-times-the-charm-kern-county-seeks-to-allow-oil-drilling-after-courts-overturn-decision-twice/}{https://www.kget.com/news/local-news/third-times-the-charm-kern-county-seeks-to-allow-oil-drilling-after-courts-overturn-decision-twice/} (accessed 31 August 2025).}
Because Kern produces over $70$ percent of oil produced within California, it
will be forced to deal with the brunt of forces that restructure how economic activity is organized along the transition away from fossil fuels. 
As a recent \textit{Economist} article put it:
``Here,
oil is not just a commodity; it is part of people's identity.
If you're not in oil, you know people who are.''\footnote{\textit{The Economist}, December 19, 2024,
  \href{https://www.economist.com/christmas-specials/2024/12/19/the-beginning-of-the-end-for-oil-in-california}{https://www.economist.com/christmas-specials/2024/12/19/the-beginning-of-the-end-for-oil-in-california}.}

At the center of the transition away from fossil fuels are two main issues.
First, there is the primary issue of jobs,
which has been the focus of much research \citep{weber_2020,raimi_etal_2023}.\footnote{\citet{weber_2020},
  for example,
  looks at job loss in the coal industry and finds that
  between 2011 and 2016,
  the number of people working in mines declined by 45 percent.
  Each lost coal-mining job is associated with a reduction in county-level income by \$100,000 annually.
  Policies that aimed to support coal-producing regions
  replaced just a fraction of this lost income \citep{weber_2020}.}
Second, there is the less appreciated issue of public finances \citep{raimi_etal_2023,raimi_etal_2024}. 
The extraction of fossil fuels in Kern County, California,
not only
provides jobs
but also
generates meaningful revenue for the local government through property taxes.
For the most part, these two literatures exist in parallel.
We link the two main issues---employment and local revenue generation---by
developing a multisector model of regional production that incorporates network-based input--output linkages.
The theory demonstrates how input--output linkages propagate and amplify the employment effects of fluctuations in crude-oil prices.
Higher crude-oil prices increase local income and demand for regional goods.
The added demand may be generated by increased purchases from either the fossil-fuel sector or the local government.
Because of input--output linkages, the net effect on overall employment may be large.
And Kern may be especially vulnerable to fewer job opportunities.
If the $58$ counties in California were ordered by per-capita personal income, from highest to lowest,
Kern would rank $54$.
Likewise,
less than $20$ percent of Kern's residents have earned a college degree.
Across the $58$ counties this ranks $48$th.\footnote{Across the United States,
  close to $39$ percent of people have earned a college degree.
  These statistics are covered in appendix \ref{sec:appendix:data-kern}.}
The constellation of issues here may highlight the need and importance of place-based policies \citep{bartik_2020jep}.

We also focus on employment outcomes as jobs are a central focus of regional development.
Within Kern County,
a major vision of the Kern Economic Development Corporation is
``advancing employment opportunities.''\footnote{``About KEDC,''
  Kern Economic Development Corporation, accessed August 31, 2025,
  \href{https://kernedc.com/about-us/}{https://kernedc.com/about-us/}.
  A similar goal is shared by A Better Bakersfield and Boundless Kern,
  often referred to as B3K.}
\citet{weber_2020} connects environmental policy and jobs
by making an analogy with trade policy.
Even though the transition away from fossil fuels may be efficient, as
nearly everyone stands to benefit a little
from reduced damages associated with higher temperatures,
these benefits are less salient than job losses.
Because oil production (like coal production) is concentrated geographically,
potential job losses are a particularly salient feature of the transition.
Ensuring employment opportunities in exposed regions
may be required to garner wider support for policy.
We therefore focus on employment.

To understand how Kern's economy depends on oil,
we propose a structural vector autoregressive model
that jointly explains the global crude-oil market and
the evolution of employment in Kern.\footnote{By focusing on the time series of employment,
  we also avoid challenging spatial econometric issues.
  For an example of these challenges, see  \citet{feyrer_mansur_sacerdote_2017}, \citet{feyrer_mansur_sacerdote_2020}, and \citet{james_smith_2020}.
  In addition,
  we lack detailed data on oil production.}
We use data
from the Bureau of Labor Statistics' Quarterly Census of Employment and Wages program.
These data cover 95 percent of US employment and are therefore
suitable to analyze local economic activity.
The monthly frequency allows
us to use what has been learned from
structural vector autoregressive models of the crude-oil market
that depend on timing assumptions \citep{kilian_2009}.

Our estimates allow us to construct counterfactual employment series,
which assess how Kern's employment would have evolved
absent the influence of oil-demand and -supply shocks.
The counterfactual employment series indicate that employment in Kern has recently benefited from favorable shocks to oil demand,
suggesting that Kern may be vulnerable to the green energy transition.
At the same time,
Kern also benefited from local shocks,
suggesting a measure of economic resilience.\footnote{\citet{modica_reggiani_2014}
  provide an overview of resilience in this context.}
Nevertheless, Kern's exposure to the green-energy transition
merits consideration of policies that aim to help Kern specifically
\citep{hanson_2023}.

To preview results briefly,
even though less than $2$ percent of jobs in Kern are in fossil-fuel-producing industries,
around $11$ percent of the variation in employment growth is explained by
the oil market.
Current employment in Kern would be roughly $5$ percent lower absent
the influence of oil-demand shocks.
The answer to the question of Kern's dependence on oil, then, suggests that there is substantial
scope for policymakers to consider tools like place-based policy
when engineering the transition away from fossil fuels.

The remainder of the paper is organized as follows.
In section \ref{sec:theory},
we develop some theoretical justifications for investigating overall employment.
Network-based production propagates and amplifies the employment effects of fluctuations in crude-oil prices.
In section \ref{sec:joint-model-global},
we describe the joint model of the global crude-oil market and
employment within Kern County.
We evaluate the dynamic effects of oil-supply and -demand shocks in the oil market in
section \ref{sec:irf-oil-block}.
We evaluate how these shocks affect regional employment in section \ref{sec:irfs-empl}.
In section \ref{sec:historical-decomp}
we study to what extent oil-market shocks explain employment growth within Kern County,
California.
Our main substantive results are presented in
section \ref{sec:counterfactual-empl},
where we focus on counterfactual employment series.
In section \ref{sec:wages},
we assess how real wages respond to unanticipated disruptions to oil supply and precautionary demand motives.
We discuss our results within the context of the literature in section \ref{sec:discussion}.
Concluding remarks are made in section \ref{sec:conclusion}.

\section{Theory}
\label{sec:theory}

The center of many people's lives is employment.
And jobs have been the focus of much research concerned with the transition away from fossil fuels, 
particularly jobs directly tied to oil and gas extraction.
\citet{raimi_etal_2023} go as far as stating
``the just-transition topic that receives the most attention from policymakers and the media can be summed up in one word: jobs'' (296).
Yet, as these authors point out, this focus ignores an equally important feature of the transition:
for state and local governments,
fossil fuels are a major source of revenue.
California collects meaningful revenue from the fossil fuels
\citep[][table 2, page 301]{raimi_etal_2023}.
More relevant for this study
is the direct revenue generated for local governments. 
As part of California's tax policy,
Kern County levies property taxes based on the assessed value of oil and gas reserves, and
``property tax revenues grow substantially with increased prices and/or production, and
can fall rapidly during a downturn'' \citep[][18]{newell_raimi_2018}.

To link the two main issues identified by \citet{raimi_etal_2023,raimi_etal_2024}---employment and local revenue generation---we
begin with a multisector general-equilibrium model of a regional economy.
The model contains many features of
a class of models used to study the macroeconomy, including
constant-returns-to-scale technology,
preferences over $n$ goods and leisure, and
firms that use intermediate goods as inputs.
This class of models, which builds on the work of \citet{long_plosser_1983}, has been used to explain large macroeconomic fluctuations
\citep{acemoglu_etal_2012,acemoglu_akcigit_kerr_2016,acemoglu_etal_2016jle,carvalho_tahbaz-salehi_2019}.
We build upon this framework by adding two features of a regional economy:
the price of oil affects 
\begin{enumerate*}[(i.)]
\item\label{feature:income} households' income and
\item\label{feature:demand} demand for sectoral output.
\end{enumerate*}
Feature \ref{feature:demand} is general enough to allow an increase in the price of oil to affect
both the oil sector and the local government's demands for inputs.
The model yields 
expressions for changes in sectoral employment resulting from changes in the price of oil.
The central idea is that a change in the price of oil that affects the crude-oil sector 
could have a large impact on overall employment if it reduces output
of other sectors connected to the crude-oil sector through input--output linkages.
This result offers a theoretical justification for investigating how overall employment in Kern County, California,
responds to changes in the global market for crude oil.

\subsection{Economic environment}

The economy is static and characterized by competitive markets.
Workers and firms take prices as given.

\subsubsection{Firms}

There are $n$ sectors indexed by $j\in\left\{ 1,\dots,n\right\}$.
A representative firm from sector $j$ has a Cobb--Douglas production function of the form
\begin{equation}
y_{j}=l_{j}^{\alpha_{j}^{l}}\prod_{i=1}^{n}x_{ji}^{a_{ji}},\label{eq:technology}
\end{equation}
where $x_{ji}$ is the quantity of goods produced by industry $i$ used as inputs by industry $j$. 
The Cobb--Douglas technology exhibits constant returns to scale;
that is, for each sector $j$,
\begin{equation}
\alpha_{j}^{l}>0,a_{ji}\geq0,\text{ and }\alpha_{j}^{l}+\sum_{i=1}^{n}a_{ji}=1.\label{eq:technology-params}
\end{equation}

\subsubsection{Market clearing}

The output of each sector is used as input for the other sectors or consumed in the final-good sector.
In addition, the crude-oil sector located in the regional economy purchases inputs from the $n$ sectors.
The purchases can be thought of as purchases made by the crude-oil sector directly or purchases made by the local government financed by property taxes paid by firms that extract fossil fuels.
We denote these purchases by the vector $\bm{z}\left(p_{o}\right)=\left(z_{1}\left(p_{o}\right),z_{2}\left(p_{o}\right),\dots,z_{n}\left(p_{o}\right)\right)\tp$, 
which makes clear that $\bm{z} \left( p_{o} \right)$ depends on the price of crude oil, $p_{o}$,
which is determined by forces of supply and demand in global markets and therefore exogenous from the perspective of the model \citep{kilian_2009}.
The main message of this paper is that this dependence amplifies and propagates oil-price shocks to the remainder of the regional economy.

In summary, the market-clearing condition for industry $j$ can be
written as
\begin{equation}
y_{j}=c_{j}+\sum_{k=1}^{n}x_{kj}+z_{j}\left(p_{o}\right),\label{eq:mkt-clearing-j}
\end{equation}
where 
$c_{j}$ is final consumption of the output produced by industry $j$ and
$z_{j}\left(p_{o}\right)$ denotes (real) purchases made by the oil sector, which depends on the price of crude oil.
This material-balance condition requires that output from sector $j$ equal total consumption plus all inputs used across industries (including industry $j$ itself) plus demand from the crude-oil sector.
As will be made apparent below, input--output linkages are captured by the terms $a_{ij}$. 

\subsubsection{Households}

A representative household is endowed with a unit of labor.
Households value consumption and leisure.
Preferences are summarized by a representative household with utility function
\begin{equation}
u\left(c_{1},\dots,c_{n},l\right)=v\left(l\right)\prod_{i=1}^{n}c_{i}^{\beta_{i}},\quad\beta_{i}\in\left(0,1\right),\label{eq:preferences}
\end{equation}
where
$l$ is the household's supply of labor,
$v$ is a decreasing and differentiable function, and
$c_{i}$ is the amount of good $i$ consumed.
The constants $\beta_{i}\geq0$ measure the share of the household's budget that is allocated to purchasing each good.
The normalization is such that $\sum_{i=}^{n}\beta_{i}=1$.
We will make use of the functional form $v\left(l\right)=\left(1-l\right)^{\lsp}$ below.

The household maximizes utility subject to the constraint that expenditure on goods must equal labor income plus income returned from the crude-oil sector:
\begin{equation}
\sum_{i=1}^{n}p_{i}c_{i}=wl+\hhincome\left(p_{o}\right),\label{eq:hh-budget-constraint}
\end{equation}
where
$w$ is the wage,
$l=\sum_{i=1}^{n}l_{i}$, and
$\hhincome\left(p_{o}\right)$ represents income earned from the crude-oil sector,
which depends on the price of crude oil.
The expression in \eqref{eq:hh-budget-constraint} is the household's budget constraint.

The added-income effect associated with a rise in the price of crude oil in the budget constraint in \eqref{eq:hh-budget-constraint} captures another important feature of the regional economy:
households' incomes may rise and fall with the price of crude oil. 

\subsubsection{Description of the competitive equilibrium}

The economic environment is specified by equations \eqref{eq:technology-params} and \eqref{eq:preferences}, along with market clearing.
Equilibrium consists of a collection of prices and quantities such that
\begin{enumerate*}[(a)]
\item\label{item:max-util} households maximizes utility subject to their budget constraints,
\item\label{item:max-profits} firms in each sector maximize profits, and  
\item\label{item:mkts-clear} all markets clear.
\end{enumerate*}
Both households and firms take prices as given.

\subsection{Main result}
\label{sec:main-result}

Starting from the competitive equilibrium,
it is straightforward to show how changes in the price of crude oil propagate to all sectors of the economy and amplifies overall changes in employment.
A detailed derivation is provided in appendix \ref{sec:app:model}.

Firms' profit motive, given the production technology in \eqref{eq:technology}, implies a representative firm's choice of input satisfies
$a_{ij}= p_{j}x_{ij} / \left( p_{i}y_{i} \right)$
[equation \eqref{eq:def-aij} in appendix \ref{sec:app:model}].
The term $p_{j}x_{ij}$ is the expenditure on input $j$, and
$a_{ij}$ represents this expenditure as a fraction of total sales. 
These are the components of the regional economy's input--output matrix $\bm{A}=\left[a_{ij}\right]$.
While the input--output matrix $\bm{A}$ will be useful to establish that prices are determined independently of the price of crude oil 
(a standard, supply-side result established in result \ref{result:prices} in appendix \ref{sec:app:model}),
it will also be useful to work with the matrix $\hat{\bm{A}}$,
whose entries are $\hat{a}_{ij} = p_{j}x_{ij} / \left( p_{j}y_{j} \right)$.
An entry of $\hat{\bm{A}}$ represents sector $i$'s expenditure on input $j$ as a fraction of sector $j$'s total sales.

In addition, we define the matrix $\hat{\bm{H}}$, along with its typical entry, as
\begin{equation}
\label{eq:Hhat}
\hat{\bm{H}} \coloneqq \left( \bm{I} - \hat{\bm{A}} \right)^{-1} = \sum\limits_{k=0}^{\infty} \hat{\bm{A}}^{k} = \bm{I} + \hat{\bm{A}} + \hat{\bm{A}}^{2} + \cdots, \quad \hat{\bm{H}} = \left[ \hat{h}_{ij} \right].
\end{equation}
And we define nominal values with tildes, so, for example, $p_{i}z_{i} = \tilde{z}_{i}$.

This notation allows us to decompose the impact of a change in the price of crude oil on sectoral employment.\footnote{For intuition about the invertibility of $\bm{I} - \hat{\bm{A}}$,
  see \citet[][32--34]{miller_blair_2022}.}
The idea is to express changes in sectoral employment, $l_{j}$,
in terms of $d\hhincome$ and $dz_{1},\dots,dz_{n}$,
where $d \hhincome \left( p_{o} \right)$ and $dz_{i} \left( p_{o} \right)$ are exogenous changes (or ``shocks'') induced by a change in the price of crude oil.\footnote{In brief,
  the technique amounts to
  totally differentiating the material-balance condition for sector $j$ with respect to $p_{o}$
  after substituting in the optimal choices of households and firms.
  Details are provided in appendix \ref{sec:app:model}.}

\begin{result}
  \label{result:main-decomposition}
  Suppose $v \left( l \right) = \left(1-l\right)^{\lsp}$.
  A change in the price of crude oil affects employment in sector $j$ through
  three separate channels:
\begin{align*}
  d\ln l_{j} &=\underbrace{\frac{\beta_{j} \times d\hhincome}{p_{j}y_{j}} \times \frac{1}{1+\lsp}}_{\text{own added-income effect}}
               + \underbrace{\frac{d\tilde{z}_{j}}{p_{j}y_{j}}}_{\text{own demand effect}} \\
  &\quad + \underbrace{\sum_{i=1}^{n}\left(\frac{\beta_{i} \times d\hhincome}{p_{i}y_{i}} \times \frac{1}{1+\lsp}+\frac{d\tilde{z}_{i}}{p_{i}y_{i}}\right)\left(\hat{h}_{ij}-\bm{1}_{i=j}\right)}_{\text{network demand effect}},
\end{align*}
where $\bm{1}_{i=j}$ is the indicator function for $i=j$.
In addition, prices are determined independently of the demand side of the economy and are therefore unaffected by changes in the price of crude oil.
And, because the wage is chosen to be the numeraire ($w=1$) and workers are paid the value of their marginal production of labor, $1=\alpha_{j}^{l}p_{j} y_{j}$,
$d \ln l_{j} = d \ln y_{j}$.
\end{result}

The own added-income effect comprises two terms:
$\beta_{j} \times d\hhincome / \left( p_{j}y_{j} \right)$ and $\left( 1-\lsp \right)^{-1}$,
where $d\hhincome$ is the change in household income associated with a change in the price of crude oil.
The first term reflects Cobb--Douglas preferences.
The representative household spends a constant fraction $\beta_{j}$ of its added income on good $j$.
This effect is moderated by the second term, $\left(1+\lsp\right)^{-1}$.
The own added-income effect is largest when there is no labor response,
which coincides with $v^{\prime}=0$ or $\lsp=0$.
When $\lsp>0$, the supply of labor is decreasing in oil income ($\hhincome$) and, therefore, the effect is lessened.

The own demand effect is the effect of direct purchases of input $j$ associated with an increase in the price of crude oil,
whether by the crude-oil sector or by the local government (financed by taxes levied on the crude-oil sector).
The value of the change is $d\tilde{z}_{j}$.

The transmission channels associated with the own added-income-effect and own demand effect are direct (the subscripts are all $j$s).
In contrast, the network effect in result \ref{result:main-decomposition}
links changes that occur in all other sectors $i \neq j$ to sector $j$.
This transmission channel reflects the regional economy's input--output linkages.
In particular,
the network demand effect comprises an indirect added-income effect and an indirect demand effect.
The intuition behind these effects are similar to the own effects.
In addition,
the own added-income effect and own demand effect propagate through the input--output linkages and are reflected in the network effect
through $\hat{h}_{jj} - 1$.\footnote{The definition of $\hat{\bm{H}}$ in \eqref{eq:Hhat} indicates that $\hat{h}_{jj} - 1$ is nonnegative.}

Fluctuations in crude-oil prices hit the fossil-fuel-extraction sector, directly.
Through network effects,
these fluctuations propagate to the entire economy and are amplified along the way.
The main message of this paper is that fluctuations in the global crude-oil market will have meaningful effects on overall employment.

\subsection{An example that illustrates how a change in the price of crude oil affects upstream sectors}
\label{sec:example-economies}

An example may best illustrate the impact of crude-oil-price changes on sectoral employment.
The example, which is motivated by the investigation in \citet{acemoglu_akcigit_kerr_2016},
clarifies the terminology used to describe ``upstream'' demand effects.
Appendix \ref{sec:app:example-economy} contains a thorough presentation of the general model and provides a thorough derivation of results.

\subsubsection{Illustrative framework}

\begin{figure}[htbp]
  \begin{subfigure}[b]{0.45\textwidth}
    \centerline{\includegraphics[width=\textwidth]{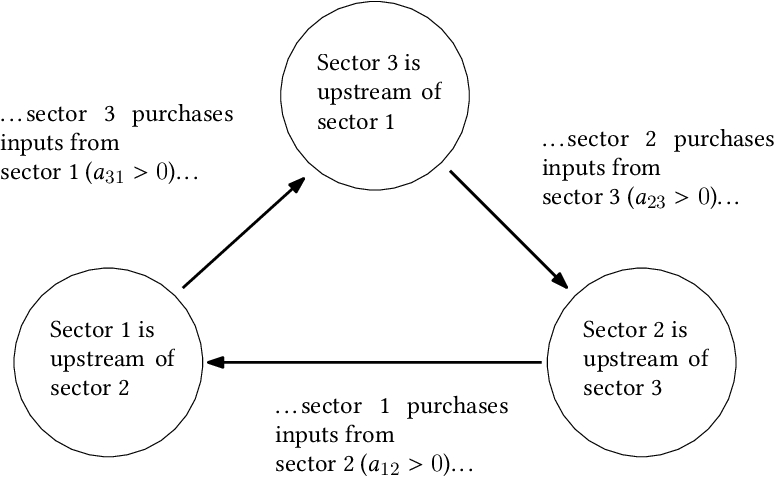}}
    \caption[]{\label{fig:ex-cycle} Complete cycle}
  \end{subfigure}
  \hfill
  \begin{subfigure}[b]{0.45\textwidth}
    \centerline{\includegraphics[width=\textwidth]{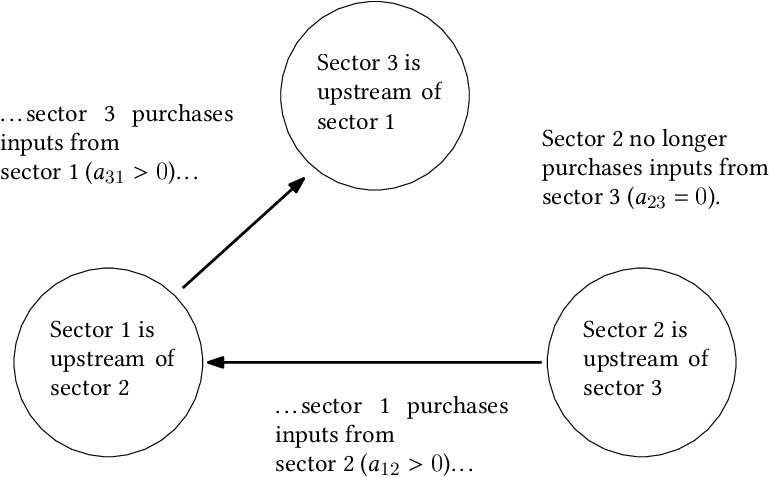}}
    \caption[]{\label{fig:ex-incomplete-cycle} Incomplete cycle}
  \end{subfigure}
  \caption[]{\label{fig:2-ex-economies} Example economies}
    \begin{figurenotes}[Notes]
      Each economy is characterized by 3 sectors.
      In panel \ref{fig:ex-cycle},
      sector 1 purchases input from sector 2 and supplies output to sector 3,
      sector 2 purchases input from sector 3 and supplies output to sector 1, and
      sector 3 purchases input from sector 1 and supplies output to sector 2.
      All sectors sell output to consumers.
      Here $a_{12}>0$, $a_{23}>0$, $a_{31}>0$, and the other $a_{ij}$s are $0$.
      In panel \ref{fig:ex-incomplete-cycle},
      the economy is characterized by a similar input--output structure, except $a_{23} = 0$.
    \end{figurenotes}
\end{figure}

Consider an economy with three sectors.
The input--output network is shown in panel \ref{fig:ex-cycle} of figure \ref{fig:2-ex-economies}.
In the economy,
sector $3$ purchases inputs from sector $1$;
sector $1$ purchases inputs from sector $2$; and
sector $2$ purchases inputs from sector $3$.
Input--output linkages exhibit a cycle. 
And it is common to say, for example, that sector $3$ is ``upstream'' of sector $1$ \citep[][S144, fn 7]{acemoglu_etal_2016jle}.

A representative firm combines a single input good with labor to produce sectoral output under constant returns to scale.
The sectoral production functions are given by
\begin{align*}
y_{1}=l_{1}^{\alpha_{1}^{l}}x_{12}^{a_{12}},\text{ }y_{2}=l_{2}^{\alpha_{2}^{l}}x_{23}^{a_{23}},\text{ and }y_{3}=l_{3}^{\alpha_{3}^{l}}x_{31}^{a_{31}}.
\end{align*}
In general,
\begin{equation}
y_{i}=e^{z_{i}}l_{i}^{\alpha_{i}^{l}}x_{ij}^{a_{ij}},\qquad\alpha_{i}^{l}+a_{ij}=1.\label{eq:ex:prod}
\end{equation}

For the sake of clarity, we assume that consumers allocate one third of their income on the expenditure of each of the three goods,
so $\beta_{i} = 1/3$.

The impact of a change in the price of crude oil on employment in sector $1$ is
\begin{equation}
  \label{eq:dl1-example}
  \begin{split}
    dl_{1} & =\frac{\alpha_{1}^{l}}{1-a_{31}a_{23}a_{12}}\left[\frac{d\hhincome}{3\left(1+\lsp\right)}\left(1+a_{31}+a_{31}a_{23}\right)\right] \\
           &\quad + \frac{\alpha_{1}^{l}}{1-a_{31}a_{23}a_{12}}\left[d\tilde{z}_{1}+a_{31}d\tilde{z}_{3}+a_{31}a_{23}d\tilde{z}_{2}\right].
  \end{split}
\end{equation}
The expression makes clear a main message of the paper:
employment in sector $1$ is influenced by features of all other sectors and effects are amplified through $\left( 1-a_{31}a_{23}a_{12} \right)^{-1} > 1$.

Consider first the case where a change in the price of crude oil has
no effect on purchases of intermediate goods, so $d\tilde{z}_{i}=0$.
This would be the case if the crude-oil sector itself did not directly
adjust its purchases of inputs or if the change in local property
taxes paid by the crude-oil sector caused no changes in local government purchases.
Then the influence of a change in the price of crude oil operates through $d\hhincome$,
the associated change in household income. 
To understand this channel, consider the coefficients associated with each sector.

The coefficient associated with sector $1$ is $1$, as shown on the
first line of equation \eqref{eq:dl1-example}.
Cobb--Douglas preferences dictate that $1/3$ the change in income, $d\hhincome$,
be allocated to purchases of good $1$.
This amount is moderated by $\left(1+\lsp\right)^{-1}$,
which reflects a household's preference for leisure.
The parameter $\alpha_{1}^{l}$ reflects the conversion of output to labor units. 

The coefficients for sectors $3$ and $2$ are $a_{31}$ and $a_{31}a_{23}$.
These values reflect the upstream position of the two industries in
input--output space, as shown in panel \ref{fig:ex-cycle} in figure \eqref{fig:2-ex-economies}.
When income changes,
a household also purchases more of good $2$ and good $3$.
Because sector $3$ is a customer of sector $1$,
using output from sector $1$ as an essential input to production,
the increase in demand for good $3$ increases demand for good $1$ by $a_{31}$,
which represents sector $3$'s input expenditure on goods produced by sector $1$ as a fraction of sales,
as indicated by \eqref{eq:def-aij}.
Likewise, sector $2$ is a customer of sector $3$.
And because sector $3$ is a customer of sector $1$,
the coefficient associated with sector $2$ is $a_{23}a_{31}$.
The effect, in other words, is ``upstream'' in the words of \citet{acemoglu_akcigit_kerr_2016} and \citet{acemoglu_etal_2016jle}.
Sectors $2$ and $3$ are both upstream of sector $1$ in input--output space. 

Next consider the case where a change in the price of crude oil affects purchases of intermediate goods,
expressed in the terms $d\tilde{z}_{1}$, $d\tilde{z}_{2}$, and $d\tilde{z}_{3}$, which is
shown on the second line of equation \eqref{eq:dl1-example}.
This channel operates through purchases made by the crude-oil sector directly in
addition to purchases made by the local government financed through property taxes paid by the crude-oil sector.

The coefficients again reflect the input--output space.
The coefficient associated with sector $1$ is $1$,
which reflects increased demand for sector $1$'s output.
The coefficients for sectors $3$ and $2$ are $a_{31}$ and $a_{31}a_{23}$,
which reflects the use of sector $1$'s output as an essential input to production. 

To see that the effects are working ``upstream'' as opposed to ``downstream,''
consider panel \ref{fig:ex-incomplete-cycle} of figure \ref{fig:2-ex-economies}.
Here sector $2$ no longer uses sector $3$'s output as an input (or $a_{23}=0$).
The expression in \eqref{eq:dl1-example} becomes
\begin{align*}
dl_{1}=\alpha_{1}^{l}\left[\frac{d\hhincome}{3\left(1+\lsp\right)}\left(1+a_{31}\right)+d\tilde{z}_{1}+a_{31}d\tilde{z}_{3}\right].
\end{align*}
The indirect effect of sector $2$ is no longer felt, even though sector $1$ is a customer of sector $2$.
Only the influence of sector $3$ is felt,
as sector $3$ is located upstream of sector $1$.\footnote{As \citet{acemoglu_akcigit_kerr_2016} explain,
the intuition for the importance of the upstream channel has to do with sectoral prices.
The constant-returns-to-scale production technology means prices are unaffected by a change in the price of crude oil.
Productivity shocks, in contrast, would affect prices, which would generate downstream effects.}

\subsection{Summary of conclusions from the illustrative framework}

Some key points about the transmission of fluctuations in crude-oil prices are worth summarizing:
\begin{itemize}
\item Fluctuations in the price of crude oil propagate throughout the regional economy through three channels:
  an own added-income effect, an own demand effect, and a network-demand effect.
  The own added-income effect and the own demand effect
  reflect the influence on the representative household's budget constraint and purchases made by the crude-oil sector or purchases financed by taxes levied on the crude-oil sector.
  The network--demand effect reflects input--output linkages.
\item Input--output linkages amplify fluctuations,
  as indicated by result \ref{result:main-decomposition} and equation \eqref{eq:dl1-example}. 
\item  An increase in the price of crude oil expands demand.
  Affected sectors need to increase their output to meet this expanded demand,
  which requires them to purchase more inputs.
  Effects propagate to sectors located upstream of affected sectors in input--output space.
\end{itemize}

We next turn to a statistical model that relates crude-oil prices to regional employment.

\section{A joint model of the global crude-oil market and employment in Kern County, California's top oil-producing region}
\label{sec:joint-model-global}

Crude oil is
an essential component of production and
a commodity traded in global markets.
In pioneering work,
\citet{kilian_2009} used these features to investigate
how oil-supply and -demand shocks affect US macroeconomic performance.
These shocks, however, may affect local economies within the United States
in different ways.
So another important reason to investigate these shocks is
understanding how oil-supply and -demand shocks affect regional economic outcomes.

We examine linkages between the global crude-oil market and regional
economic outcomes by specifying and estimating a structural VAR model.
The proposed VAR model
relies on insights made by \citet{kilian_2009} about the crude-oil market and
incorporates the widely held view that jobs are a major focus of regional development
\citep[see, for example,][]{bartik_2020,weber_2020}.
The proposed VAR model jointly explains the crude-oil market and regional employment.

In trying to understand local employment dynamics,
we focus on the global crude-oil market for a few reasons.
The primary reason is our lack of readily available microlevel data at a monthly frequency,
which is part of our identification strategy.
California is also a net importer of crude oil.
This allows us to use US refiners' acquisition cost of imported crude oil as the relevant price.\footnote{``Annual Oil Supply Sources to California Refineries,''
  California Energy Commission,
  accessed February 23, 2026, \href{https://www.energy.ca.gov/data-reports/energy-almanac/californias-petroleum-market/annual-oil-supply-sources-california}{\nolinkurl{https://www.energy.ca.gov/data-reports/energy-almanac/californias-petroleum-market/annual-oil-supply-sources-california}}.
  We thank an anonymous referee for help with this discussion.}
Finally,
the direct effect that oil-supply and -demand shocks have on employment is likely small
(less than $2$ percent of workers in Kern are employed in the oil sector).
Yet, the indirect effects may be large.
This hypothesis is motivated by the theoretical framework presented in section \ref{sec:theory},
which shows how shocks propagate throughout a regional economy and amplify the response of employment. 
In general, fluctuations in the forces that determine supply and demand for crude oil
will affect employment dynamics in different ways.
For example,
as we will document empirically below,
an increase in the precautionary demand for crude oil will cause an immediate and persistent increase in the price of crude oil;
whereas,
an unanticipated disruption to crude-oil production will cause a delayed increase.
Capturing these dynamics requires that we adopt the pioneering methodology of \citet{kilian_2009},
which allows us to structurally decompose changes in employment into three separate components:
crude-oil supply shocks, shocks to the global demand for all industrial commodities, and oil-specific demand shocks.

In brief:
our proposed model
captures the net, general-equilibrium effects and
allows us to isolate how different factors that explain the global crude-oil market  have affected regional employment growth.

\subsection{Key determinants} 

We build on \citeauthor{kilian_2009}'s \citeyearpar{kilian_2009}
pioneering model of the global crude-oil market.
The model comprises three key determinants of the real price of crude oil.
The first is an \textbf{oil-supply shock} that reflects unexpected disruptions to
crude-oil availability.
The two other determinants affect demand.
The \textbf{aggregate-demand shock} allows global real economic activity to shift the
demand for crude oil,
consistent with the consensus that a global business cycle affects demand
for commodities traded in large markets.
The \textbf{precautionary-demand shock} allows demand to shift
when future availability of oil is a concern.
\citet{alquist_kilian_2010} demonstrate the importance of this component.
Notably,
this component predicts that the real price of oil will ``overshoot''
in response to a precautionary-demand shock,
which is consistent with the empirical evidence we will share below.\footnote{\citet{knittel_pindyck_2016} provide a general overview of related ideas.}

We link the crude-oil market to regional employment outcomes through a single channel.
This channel reflects shocks to Kern's employment growth
not driven by global crude-oil demand or supply.
Many different types of shocks affect regional employment through this channel.
We do not attempt the monumental challenge of classifying each one, because
we are interested in how the crude-oil market affects employment in Kern County, California.
The identifying assumption that allows us to quantify the regional impact of the global crude-oil market
is a timing restriction.
We maintain that shocks to Kern's employment growth do not affect the global crude-oil market within the month.
This assumption is consistent with evidence that \citet{kilian_vega_2011} provide on
energy prices and US macroeconomic aggregates.

In summary,
the structural VAR model we propose jointly explains four variables of interest:
\begin{enumerate*}[(i)]
\item\label{item:3} the log difference of global crude-oil production,
\item\label{item:4} a measure of global real economic activity,
\item\label{item:5} the log of the real price of oil, and
\item\label{item:6} the log difference of employment in California's top oil-producing region,
  Kern County.
\end{enumerate*}
Data on global crude-oil production are available from the US Energy Information Administration.
While there are numerous ways to measure global real economic activity,
we use the measure proposed by \citet{kilian_2009} and updated by \citet{kilian_2019}.
This measure is based on dry-cargo shipping rates.
It serves as a proxy for the global movement of industrial commodities.
The series is readily available through FRED (Federal Reserve Economic Data) and 
its merits are discussed by \citet{kilian_zhou_2018}.\footnote{Federal Reserve Bank of Dallas, Index of Global Real Economic Activity [IGREA], retrieved from FRED, Federal Reserve Bank of St. Louis; \href{https://fred.stlouisfed.org/series/IGREA}{https://fred.stlouisfed.org/series/IGREA}.}
The real price of oil is constructed by
taking the series for US refiners' acquisition cost of imported crude oil and
deflating the series using the Consumer Price Index for All Urban Consumers.\footnote{U.S. Bureau of Labor Statistics, Consumer Price Index for All Urban Consumers: All Items in U.S. City Average [CPIAUCSL], retrieved from FRED, Federal Reserve Bank of St. Louis;
  \href{https://fred.stlouisfed.org/series/CPIAUCSL}{https://fred.stlouisfed.org/series/CPIAUCSL}.}

We use employment data from the
\href{https://www.bls.gov/cew/}{Quarterly Census of Employment and Wages} program to measure employment.
These data begin in January 1990.
And they allow us 
not only
to measure county-level labor-market conditions 
but also
to provide evidence about employment in fossil-fuel producing sectors. 

The QCEW program classifies establishments to industries according to the
North American Industry Classification System (NAICS),
which allows us to look at employment in detailed industries.
A notable finding from these data is how few workers are directly employed in industries
that extract oil and gas.
Over the period January 1990 through December 2024,
less than three percent of workers directly extracted oil and gas.
And the share of workers in this sector is trending down.
More workers are employed in industries that support oil and gas extraction.
But these data also count support activities for mining.\footnote{Figure \ref{fig:empl-share-oil} in the appendix shows these data.}
Currently, the industry associated with the two-digit NAICS code for mining and oil-and-gas extraction
employs less than $2$ percent of workers in Kern County, California.

These statistics contradict the predominant view held in Kern
that oil extraction is tantamount to economic opportunity.
The view about
the wider role that oil plays likely has to do with
property taxes paid by oil companies,
donations afforded by money made from oil, and
networks that link different sectors of the economy.\footnote{For example, in the fiscal year that ended on June 30, 2024,
  Chevron paid the most property taxes in Kern County
  (\href{https://www.auditor.co.kern.ca.us/cafr/24CAFR.pdf}{https://www.auditor.co.kern.ca.us/cafr/24CAFR.pdf}) (accessed 30 August 2025).}
We are interested in
this narrative and
the general-equilibrium effects of oil-market shocks.
As such, we use the log difference of aggregate employment in Kern County, California.
This measure includes workers classified as employed by the QCEW program
across all industries and across government and privately operated establishments.

\subsection{A structural model that decomposes regional employment growth}

The proposed VAR model jointly explains the variables collected in the vector 
\begin{align*}
  y_{t} = \left(
  \Delta \; \text{oil production},
  \text{real activity},
  \text{real price of oil},
  \Delta \; \text{local employment}
  \right)^{\prime}.
\end{align*}
The structural VAR representation is
\begin{equation}
\label{eq:svar}
  B_0 y_t = \beta + B_1 y_{t-1} + \cdots + B_{12} y_{t-12} + w_t,
\end{equation}
where 
the $B_{i}$ are $4 \times 4$ parameter matrices and
$w_t \sim \left( 0, \Sigma_{w}\right)$; that is,
$w_t$ denotes a vector of serially and mutually uncorrelated innovations.
The covariance matrix $\Sigma_{w}$ is diagonal.  
The corresponding reduced-form model is
\begin{equation}
\label{eq:ols}
y_t = \alpha + A_1 y_{t-1} + \cdots + A_{12} y_{t-12} + u_t,
\end{equation}
where $A_i = B_0^{-1}B_i$ and 
$u_t = B_0^{-1} w_t \sim \left( 0 , \Sigma_{u} \right)$ is a martingale-difference
sequence with positive-definite covariance matrix
$\Sigma_{u} = \left( B_0^{-1} \right) \Sigma_{w} \left( B_0^{-1} \right)^{\prime}$.  
Going between the structural model in
\eqref{eq:svar} and the reduced-form model in \eqref{eq:ols} requires
an estimate of $B_0^{-1}$.  
Short-run identifying restrictions, discussed below, make estimating this term straightforward
using a Cholesky decomposition.

The identifying restrictions allow us to interpret correlations between
the components of the global oil market and
local labor-market conditions
causally.
We need to be precise, however, about the timing of various shocks. 
We posit that \citeauthor{kilian_2009}'s \citeyearpar{kilian_2009} identifying assumptions hold.
We discuss these assumptions next.
In addition,
we posit that employment growth in Kern---within the month---does not
affect 
global oil production, 
global bulk dry-cargo shipping rates, and
the real price of oil.
Support for the
assumption that county-level employment growth
does not affect the real price of oil is provided by
\citet{kilian_vega_2011},
who 
look at how macroeconomic news
(the difference between survey expectations and the
realizations of macroeconomic aggregates)
affects the real price of oil.
They find no meaningful evidence that news
affects oil prices
(unlike stock prices, bond prices, and exchange rates),
consistent with the identifying assumption that there is ``no feedback from US macroeconomic aggregates to monthly innovations in energy price'' (660).

The timing assumptions imply a recursive structure for $B_0^{-1}$.
The reduced-form error $u_t$ can be decomposed as
\begin{equation}
\label{eq:identification}
u_t 
\equiv 
\begin{bmatrix*}[l]
u_t^{\Delta \text{ oil production}} \\
u_t^{\text{real activity}} \\
u_t^{\text{real price of oil}}  \\
u_t^{\Delta \text{ Kern employment}} 
\end{bmatrix*}
= 
\begin{bmatrix}
b_{11} & 0      & 0     & 0 \\
b_{21} & b_{22} & 0     & 0 \\
b_{31} & b_{32} & b_{33} & 0 \\
b_{41} & b_{42} & b_{43}  & b_{44}
\end{bmatrix}
\begin{bmatrix*}[l]
w_t^{\text{oil supply}} \\
w_t^{\text{aggregate demand}} \\
w_t^{\text{oil-specific demand}} \\
w_t^{\text{regional-employment demand}} 
\end{bmatrix*}.
\end{equation}
The system in \eqref{eq:identification}
can be interpreted as $4$ equations.

The first equation indicates that the stochastic short-run supply curve is vertical: 
oil production does not respond to innovations in demand within the month.
The maintained assumption is that adjusting supply is prohibitively costly,
so that short-run demand innovations are postulated to have a negligible effect on supply within the month. 
This assumption is consistent with microeconomic evidence documented by
\citet{anderson_kellogg_salant_2018}.
They show how drilling immediately responds to 
increases in oil prices but
production responds only with delay,
suggesting that the monthly, stochastic, short-run supply curve is vertical.

In contrast,
the short-run demand curve is downward sloping.
There are two demand shocks and
each is identified by delay restrictions.
The short-run demand curve can be shifted by 
innovations to aggregate demand and innovations to oil-specific demand
(such as precautionary-demand innovations).
The delay restriction is such that oil-specific demand shocks do not affect
global real economic activity as measured by fluctuations in shipping rates relative to trend.
These assumptions are shown in equations two and three of the system listed in \eqref{eq:identification}.
Finally, the fourth equation of the system in \eqref{eq:identification}
indicates that innovations to regional employment growth do not affect the real price of oil within the month \citep{kilian_vega_2011}.

The reduced-form model in \eqref{eq:ols} includes $12$ lags and
is estimated using least squares.\footnote{The choice of lag order
represents a compromise between
capturing dynamics estimated by \citet{kilian_2009} using $24$ lags and
having enough data to estimate the larger system using data that begin in January $1990$.}
A Cholesky decomposition is used to decompose the estimate of
$\widehat{\Sigma}_{u}$ to produce an estimate $B_{0}^{-1}$.
Inference is based on a recursive-design residual-block-bootstrap procedure
proposed by \citet{bruggemann_jentsch_trenkler_2016} and
discussed by \citet{kilian_lutkepohl_2017}.
The bootstrap uses $2,000$ replications and
a block length of $24$ months.

\section{Empirical results}
\label{sec:empirical-results}

\subsection{How demand and supply shocks in the global crude-oil market affect global oil production, real economic activity, and the real price of oil}
\label{sec:irf-oil-block}

The structural impulse responses define
the responses of elements of $y_{t}$ to a one-time impulse in $w_{t}$.
Impulse response estimates are shown for a horizon of $15$ months.
The innovations are normalized so that each innovation
will tend to raise the price of oil (negative supply shocks and positive demand shocks).
Because oil production and employment growth are entered as log differences,
we report the cumulative responses of these two variables.

\begin{figure}[htbp]
  \centering
  \resizebox{\textwidth}{!}{
  \input{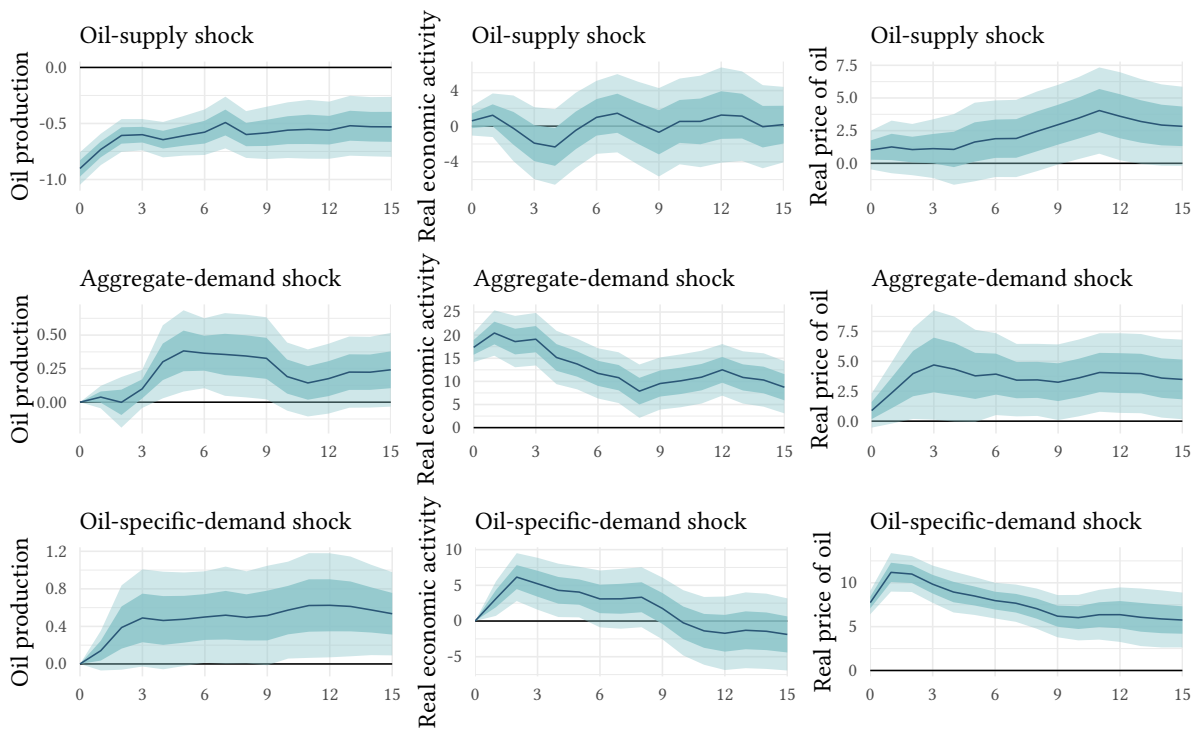}}
\caption{Responses to one-standard-deviation structural shocks}
\label{fig:irf-oil-block}
\begin{figurenotes}[Notes]
  Estimates based on the model in \eqref{eq:svar}.
  The panels show point estimates with one- and two-standard-error bands.
  Confidence intervals were constructed using a moving block bootstrap method designed by \citet{bruggemann_jentsch_trenkler_2016}.
\end{figurenotes}
\end{figure}

Figure \ref{fig:irf-oil-block} shows the
responses of global oil production, real economic activity, and the real price of oil.
The estimated responses indicate that
\citeauthor{kilian_2009}'s \citeyearpar{kilian_2009} qualitative conclusions
about the global crude-oil market hold.\footnote{To be clear,
  \citet{kilian_2009} considers only the global crude-oil market and
  estimates the structural model over a different period.
  See \citet{ryan_michieka_2025} for a replication of
  \citeauthor{kilian_2009}'s \citeyearpar{kilian_2009} results
  that includes updating the sample used in the original analysis.}

Looking at the first row of figure \ref{fig:irf-oil-block},
an unexpected disruption to oil supply
causes an immediate drop in oil production.
The effect is persistent.
And while the disruption is steadily reversed after the impact of the shock,
after 15 months, oil supply is still lower by about half of the immediate impact.
The second panel shows that an unexpected fall in production
has little effect on real economic activity.
This result provides more evidence for the insight made by \citet{kilian_2009}.
Oil-supply disruptions are less an explanation for sluggish real economic activity
than widely thought.
(It is the fear of future unavailability of oil that drives higher prices,
which hampers growth.)
The third panel of the first row of figure \ref{fig:irf-oil-block}
shows that a supply disruption
causes the price of oil to rise on impact, but
the effect is only marginally statistically significant based
on the one-standard-error bands.
The persistence of the shock, however, causes the real price of oil
to rise steadily over the next $11$ months.
By this point, the effect is statistically significant based on the
two-standard-error bands.\footnote{The estimates in \citet{kilian_2009}
  provide less evidence for this effect,
  which may well reflect different sample periods.}

The second row of figure \ref{fig:irf-oil-block}
shows the responses to an unexpected expansion of real economic activity.
The first panel shows that
it takes around $3$ months for oil supply
to respond to an increase in real economic activity.
But after $3$ months, the effect is statistically significant.
Even $15$ months later oil production is higher,
although the effect is significant based on the one-standard-error bands only.
The second panel
shows that an unanticipated rise in aggregate demand causes a
large, significant jump in global real economic activity.
The effect subsides after impact but is persistent.
Real economic activity is significantly higher over a year later.
The third panel
shows that a surprise expansion of the global business cycle causes
the real price of oil to rise.
The real price of oil is higher on impact and continues to move higher over the next $3$ months.
The effect is arguably statistically significant using the two-standard-error bands.
After $3$ months, the price declines but is still elevated $15$ months after impact.

The third row of figure \ref{fig:irf-oil-block} 
shows the responses to a surprise increase in precautionary demand for oil.
The first panel shows that
the unanticipated increase in demand causes supply to slowly expand after impact.
Supply expands but the expansion is significant based only on
the one-standard-error bands.
After $9$ months the expansion is significant based on
the two-standard-error bands.
After $15$ months supply is higher,
suggesting that higher oil prices are accommodated by expanded supply.
The second panel shows that the unanticipated rise in
oil-specific demand is associated with an increase in economic activity.
Producers may benefit from expanded oil production, for example.
But after around $9$ months,
the higher prices push real economic activity lower.
Finally,
the third panel shows the precautionary-demand motive.
The real price of oil jumps upward immediately and continues to rise for a couple months.
This overshooting effect is predicted by models of precautionary demand,
where there is a convenience factor associated with having oil on hand.
Higher prices encourage firms to demand more oil out of fear of running out in the future.

The impulse response functions presented in figure \ref{fig:irf-oil-block}
are entirely consistent with what is found in the literature.
\citet{zhou_2020},
using a second-generation model of the global crude-oil market
based on sign restrictions and global crude-oil inventories,
for example, reports results that are remarkably similar to our results.\footnote{\citet{zhou_2020} provides an excellent entry point into this literature.
See also \citet{kilian_2019}.}
In summary,
the structural model based on \citeauthor{kilian_2009}'s \citeyearpar{kilian_2009}
pioneering work,
produces dynamics that are consistent with what is known  in the literature
about the global crude-oil market.

\subsection{How demand and supply shocks in the global crude-oil market affect employment in Kern}
\label{sec:irfs-empl}

\begin{figure}[htbp]
  \centering
  \resizebox{0.6\textwidth}{!}{
  \input{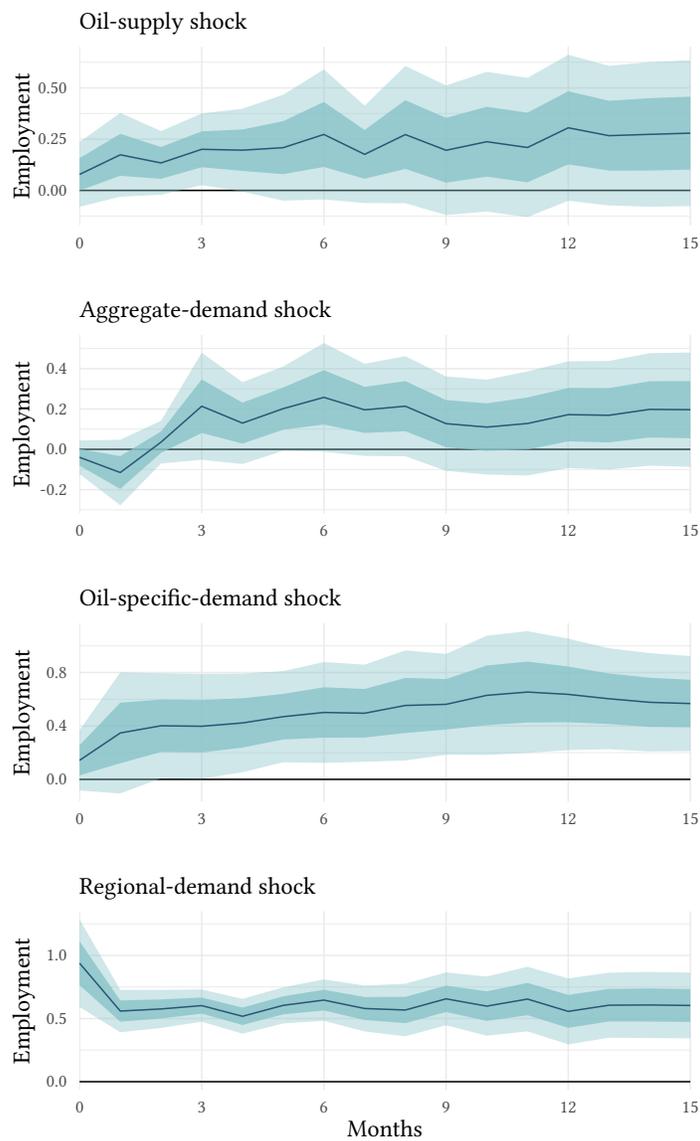}}
\caption{Cumulative employment responses to one-standard-deviation structural shocks in Kern County, California}
\label{fig:irf-empl-block}
\begin{figurenotes}[Notes]
  Estimates based on the model in \eqref{eq:svar}.
  The panels show point estimates with one- and two-standard-error bands.
  Confidence intervals were constructed using a moving block bootstrap method designed by \citet{bruggemann_jentsch_trenkler_2016}.
\end{figurenotes}
\end{figure}

Given that we have captured realistic oil-market dynamics,
how do oil-demand and oil-supply shocks affect employment
in California's top oil-producing region?
Figure \ref{fig:irf-empl-block} shows the cumulative response of employment
to one-standard-deviation structural shocks associated with the global crude-oil market.

An unanticipated disruption to oil supply causes employment to
increase initially and expand generally over the next $15$ months,
although the effect is only significant based on the one-standard-error bands.
From the perspective of Kern,
a supply disruption that causes the real price of oil to rise
is like a windfall boost to regional income.
The effect evidently outweighs the negative effects on economic activity associated with
higher oil prices.
Which suggests that there may be substantial linkages between
oil prices and the regional economy in Kern.

A surprise increase in global real economic activity expands employment opportunities.
This can be seen in the second panel of figure \ref{fig:irf-empl-block}.
There is little effect on impact, but
after $3$ months employment is higher based on the one-standard-error bands.
There is some statistical evidence that employment is higher $15$ months later.

In contrast,
a precautionary demand shock that pushes the real price of oil higher
leads to a significant expansion of employment opportunities.
Employment is higher $3$ months after impact
based on the two-standard-error bands and remains higher
$15$ months after the initial shock.
The windfall boost to regional income outweighs the negative economic effects of
higher oil prices.
In addition,
the result repeats an insight made by \citet{kilian_2009}.
While both the negative supply shock and the positive demand shock
raise the real price of oil,
only the increase in price associated with the precautionary demand shock
leads unambiguously to expanded employment.

Finally,
an unanticipated shock to employment growth causes employment to jump up on impact.
The expansion is partly reversed in the first month,
but employment remains persistently higher $15$ months after impact.

A main substantive result is that employment opportunities depend on
the global crude-oil market,
suggesting that there may be meaningful network effects between sectors.
This result is the main topic of subsequent sections.

\subsection{What explains the fluctuations in employment growth?}
\label{sec:historical-decomp}

\begin{figure}[htbp]
  \centering
  \resizebox{0.8\textwidth}{!}{
  \input{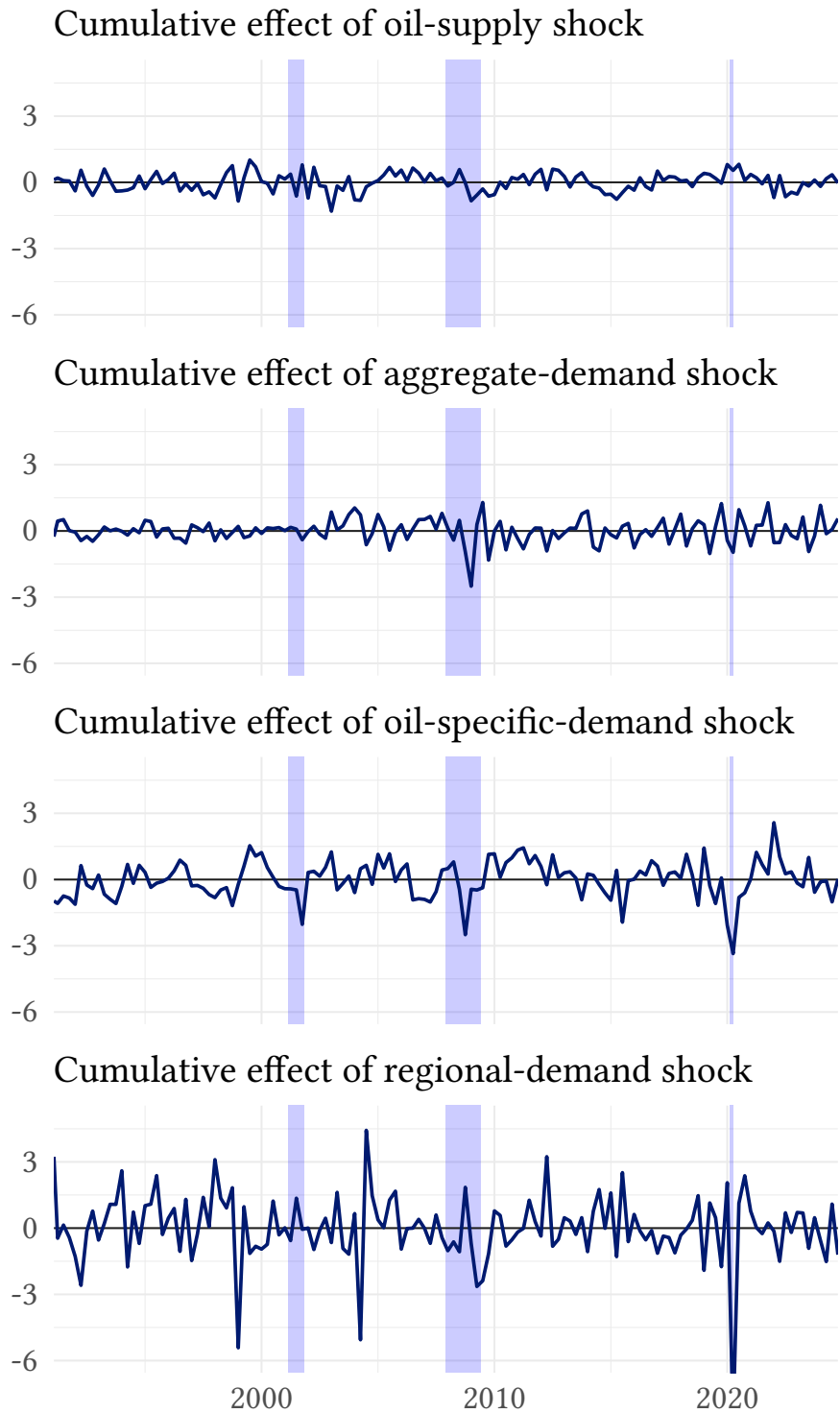}}
\caption{Historical decomposition of employment growth in Kern County, California,
January $1995$ through December $2024$}
\label{fig:historical-decomp}
\begin{figurenotes}[Notes]
  Estimates derived from the model in \eqref{eq:svar}.
  Shaded areas indicate US recessions.
\end{figurenotes}
\end{figure}

The cumulative effect that each of the structural oil-supply and oil-demand shocks
has had on employment growth since January $1995$
is shown in figure \ref{fig:historical-decomp}.
The historical decomposition is based on the estimate of the structural model in \eqref{eq:svar}.
The demand and supply shocks have been aggregated to a quarterly frequency
by computing the quarterly average of monthly values.
To interpret the historical decomposition reported in figure \ref{fig:historical-decomp},
the quarterly growth rate of employment 
is well approximated by the sum of the four lines across the $4$ panels.

The top panel of figure \ref{fig:historical-decomp}
shows that crude-oil supply shocks have had a comparatively small
effect of employment growth.
This finding is broadly consistent with related evidence
reported by \citet{kilian_2009}.
Shortfalls in oil supply have contributed little to high oil prices.
Likewise, in Kern,
other than Kern-specific shocks,
the aggregate-demand and oil-specific-demand shocks explain the
majority of fluctuations in employment growth.
The second panel indicates that
before $2005$
aggregate-demand shocks explain little of observed fluctuations in employment growth and
around the Great Recession
aggregate-demand shocks explain more of the observed fluctuations in employment growth.
The third panel indicates that
oil-specific-demand shocks
have a more persistent effect on growth rates
(the line exhibits fewer ups and downs),
especially around recessions.
Perhaps not surprisingly,
the Kern-specific-demand shocks explain the vast majority
of fluctuations in employment growth rates.

Figure \ref{fig:historical-decomp} also echoes
\citeauthor{kilian_2009}'s \citeyearpar{kilian_2009} finding
that ``not all oil-price shocks are alike.''
Oil-supply shocks have explained much less of the observed fluctuations in employment
growth
compared to oil-demand shocks.

\subsection{Historical counterfactuals indicate Kern's dependence on oil and signify resilience}
\label{sec:counterfactual-empl}

The results so far document that
employment opportunities in California's top oil-producing region respond meaningfully to fluctuations in the oil market.
To further understand 
this relationship,
we propose constructing counterfactuals,
using the techniques described by \citet{kilian_lee_2014}.
We investigate two separate periods:
the period January $1997$ through December $2009$
(over which Kern's employment rose $17$ percent, from $223,913$ to $261,258$) and
the period January $2010$ to December $2024$
(over which Kern's employment rose $28$ percent, from $266,426$ to $341,210$).
The counterfactual series
indicate how employment would have evolved 
had one of the structural shocks been replaced everywhere by zero.
So we are able to assess the importance of each shock over the two periods.

The starting point is the historical decomposition that undergirds figure \ref{fig:historical-decomp}.
The decomposition is based on the Wold decomposition theorem and moving-average representation of the system.\footnote{\citet{kilian_lutkepohl_2017} provide an excellent discussion.}
Let $t \in \left\{ 1,\dots,t \right\}$ index time in the sample.
Then the $4 \times 1$ vector $y_{t}$ can be written as
\begin{equation}
\label{eq:hist-decomp}
y_{t} = \sum\limits_{i=0}^{\infty} \Theta_{i}w_{t-i} \approx \sum\limits_{i=0}^{t-1} \Theta_{i}w_{t-i},
\end{equation}
where $\Theta_{i}$ denotes a $4 \times 4$ matrix of structural responses at lag $i=0,1,2,\dots$, and
$w_{t}$ denotes the structural shocks.
The approximation is based on the idea that the moving-average coefficients die out.
Terms in the far past have little effect on $y_{t}$.
Using the estimates of the structural model in \eqref{eq:svar},
\begin{equation}
\label{eq:2}
\hat{y}_{t} \approx \sum\limits_{i=0}^{t-1} \widehat{\Theta}_{i}\hat{w}_{t-i}.
\end{equation}

The term $y_{4t}$ (the fourth component of the vector $y_{t}$) is the growth rate of employment in period $t$.
The estimate $\hat{y}_{4t}$ can be written as $\hat{y}_{4t} = \sum\limits_{j=1}^{4} \hat{y}_{4t}^{\left( j \right)}$,
where $\hat{y}_{4t}^{\left( j \right)}$ represents the cumulative effect of structural shock $j$ on $y_{4t}$.

Following \citet{kilian_lee_2014},
this allows us to construct $y_{4t} - \hat{y}_{4t}^{\left( j \right)}$,
which tells us how employment would have evolved absent the influence of structural shock $j$.
To make the counterfactual as concrete as possible,
we construct the counterfactual in terms of employment levels.
We do this by starting with employment in period $t_{\circ}$ and using $y_{4t_{\circ}} - \hat{y}_{4t_{\circ}}^{\left( j \right)}$ to construct employment in period $t_{\circ} + 1$.
Then using $y_{4t_{\circ}+1} - \hat{y}_{4t_{\circ}+1}^{\left( j \right)}$ to construct employment in period $t_{\circ} + 2$.
Because the moving-average representation is based on a zero-mean process,
we also add in the average growth rate of employment over the period in question.

We consider two periods.
The first falls between January 1997 and December 2009.
The second falls between January 2010 and December 2024.
The later period corresponds with the rise in fracking across the United States
(although, extraction in Kern was noticeably unaffected by the fracking boom).
For each period, we separately remove the contribution of each shock. 

\begin{figure}[htbp]
  \centering
  \resizebox{\textwidth}{!}{
  \input{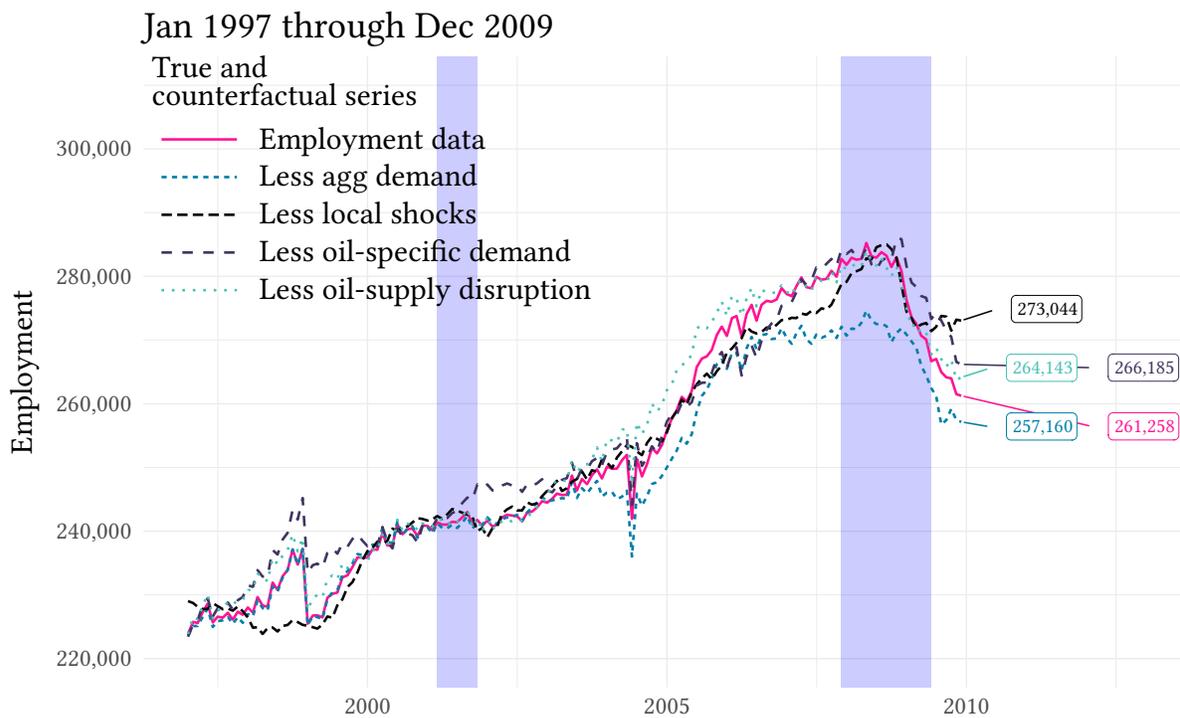}}
\caption{Counterfactuals for employment, January $1997$ through December $2009$}
\label{fig:historical-counterfactual-early}
\begin{figurenotes}[Notes]
  Actual employment is highlighted in pink.
  The counterfactuals show the evolution of employment in the absence of the listed structural shock.
  If the counterfactual series falls below actual employment,
  then the shock in question boosted employment.
  Shaded areas indicate US recessions.
\end{figurenotes}
\end{figure}

Figure \ref{fig:historical-counterfactual-early} shows the counterfactual employment series before $2010$.
Actual employment data are depicted in pink.
Early in this period,
the global crude-oil market has little influence on employment in Kern County, California.
The influence grows over time, however.

Table \ref{tab:counterfactual-empl-early} provides some numbers to understand this pattern.
In January 2000, employment is $235,843$ and employment absent any of the four shocks is nearly the same.
In contrast,
in January 2005, which is shown in figure \ref{fig:historical-counterfactual-early}, unanticipated oil-supply disruptions lowered employment.
Absent this shock
employment would have been around $2.5$ percent lower.

By December 2009, 
employment in California's top oil-producing region was $261,258$.
This number is reported on the right side of the figure \ref{fig:historical-counterfactual-early}.
The series that shows counterfactual employment less the influence of local shocks,
indicates that employment would have been $273,044$ if the Kern-specific shock to employment was absent.
Put differently,
the local shock lowered employment by about $4.5$ percent.
This result is consistent with labor-market dynamics after the Great Recession.
The labor market continued to weaken after the Great Recession was declared over.

\begin{table}[h]
  \centering
  \caption{Counterfactual employment values, January $1997$ to December $2009$}
  \label{tab:counterfactual-empl-early}
  \resizebox{\textwidth}{!}{
    
\begin{tabular}{lcrcrcr}
\toprule
\multicolumn{1}{c}{ } & \multicolumn{2}{c}{Jan 2000} & \multicolumn{2}{c}{Jan 2005} & \multicolumn{2}{c}{Dec 2009} \\
\cmidrule(l{3pt}r{3pt}){2-3} \cmidrule(l{3pt}r{3pt}){4-5} \cmidrule(l{3pt}r{3pt}){6-7}
Counterfactual & Employment & Percent & Employment & Percent & Employment & Percent\\
\midrule
\textbf{Employment data} & 235,843 & 0.0 & 255,887 & 0.0 & 261,258 & 0.0\\
Less oil-supply disruption & 235,324 & $-$0.2 & \textcolor{blue}{262,176} & \textcolor{blue}{2.5} & 264,143 & 1.1\\
Less agg demand & 236,346 & 0.2 & 250,036 & $-$2.3 & 257,160 & $-$1.6\\
Less oil-specific demand & \textcolor{blue}{237,650} & \textcolor{blue}{0.8} & 257,223 & 0.5 & 266,185 & 1.9\\
Less local shocks & 236,899 & 0.4 & 255,560 & $-$0.1 & \textcolor{blue}{273,044} & \textcolor{blue}{4.5}\\
\bottomrule
\end{tabular}
}

  \vspace{0.5em}
  
  \begin{tablenotesminipage}[Notes]
    Actual levels of employment are shown in the row ``\textbf{Employment data}.''
    The values are derived from the statistical model in \eqref{eq:svar} and constructed to
    match actual employment growth over the period January $1997$ to December $2009$.
    If the counterfactual series falls below actual employment,
    then the shock in question boosted employment.
    In each of the three periods, the shock with the largest influence on employment is highlighted in blue.
    The values in the table are shown in figure \ref{fig:historical-counterfactual-early}.
  \end{tablenotesminipage}
\end{table}

Overall,
figure \ref{fig:historical-counterfactual-early} shows that none of the $4$ structural shocks played a predominant role in explaining employment over this period,
which contrasts with the later period.
Nevertheless, two features of the figure stand out.
First, oil-specific-demand shocks had little influence.
Second,
starting around $2003$,
the aggregate-demand shock raised employment and
the oil-supply shock lowered employment.
By January $2005$,
the aggregate-demand shock raised employment by $2.3$ percent and
the oil-supply shock lowered employment by $2.5$ percent,
as reported in the fifth column of table \ref{tab:counterfactual-empl-early}.
Even though oil-supply shocks explain little of the ups and downs of employment growth,
visible in the historical decomposition in figure \ref{fig:historical-decomp},
the series of supply shocks was able to push employment lower in Kern around $2005$.

\begin{figure}[htbp]
  \centering
  \resizebox{\textwidth}{!}{
  \input{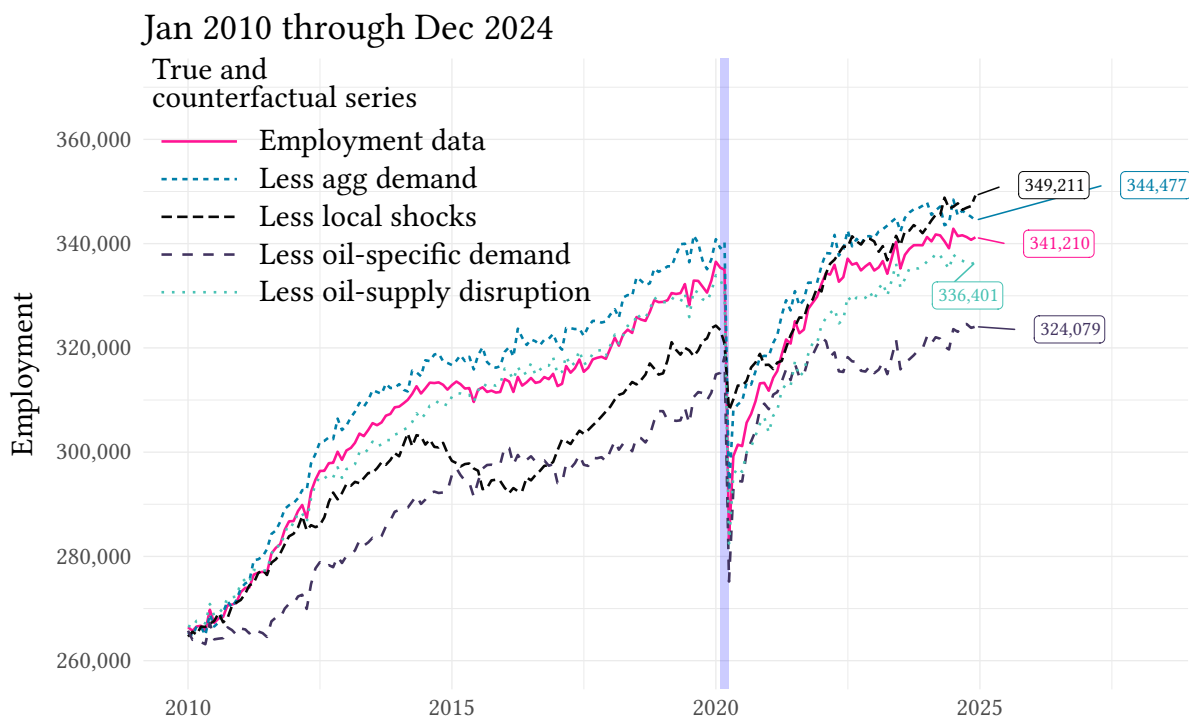}}
\caption{Counterfactuals for employment, January 2010 through December 2024}
\label{fig:historical-counterfactual-late}
\begin{figurenotes}[Notes]
  Actual employment is highlighted in pink.
  The counterfactuals show the evolution of employment in the absence of the listed structural shock.
  If the counterfactual series falls below actual employment,
  then the shock in question boosted employment.
  In each of the three periods, the shock with the largest influence on employment is highlighted in blue.  
  Shaded areas indicate US recessions.
\end{figurenotes}
\end{figure}

Figure \ref{fig:historical-counterfactual-late} shows the importance of particular shocks in explaining different historical episodes;
in particular,
throughout the period, oil-specific-demand shocks boosted employment.
By the end of $2024$,
employment in Kern was $341,210$.
Absent the influence of oil-specific-demand shocks, 
employment would have been $324,079$, a $5$ percent drop.
This statistic is reported in the seventh column of table \ref{tab:counterfactual-empl-late},
which also reports the positive influence on employment of the oil-specific-demand shock in January 2020 and January 2015.
In January 2020, on the eve of the Covid-19 recession,
employment would have been $6.4$ percent lower absent oil-specific-demand shocks.
Similarly, in January 2015,
employment would have been $5.5$ percent lower absent oil-specific-demand shocks.

The influence of oil-specific-demand shocks clearly stands out.
In figure \ref{fig:historical-counterfactual-late},
the broken, purple line associated with oil-specific-demand shocks lies below the solid, pink line associated with actual employment, uniformly, and
table \ref{tab:counterfactual-empl-late} highlights this pattern of influence in blue.
This likely has to do with fracking, even though fracking has been much, much less influential in California compared to other states
like Texas, New Mexico, and North Dakota
\citep[see also][]{hausman_kellogg_2015,bartik_etal_2019}.
Regardless,
throughout the period,
oil-specific-demand shocks boosted employment opportunities in Kern.

\begin{table}[h]
  \centering
  \caption{Counterfactual employment values, January 2010 to December 2024}
  \label{tab:counterfactual-empl-late}
  \resizebox{\textwidth}{!}{
    
\begin{tabular}{lcrcrcr}
\toprule
\multicolumn{1}{c}{ } & \multicolumn{2}{c}{Jan 2015} & \multicolumn{2}{c}{Jan 2020} & \multicolumn{2}{c}{Dec 2024} \\
\cmidrule(l{3pt}r{3pt}){2-3} \cmidrule(l{3pt}r{3pt}){4-5} \cmidrule(l{3pt}r{3pt}){6-7}
Counterfactual & Employment & Percent & Employment & Percent & Employment & Percent\\
\midrule
\textbf{Employment data} & 312,867 & 0.0 & 336,524 & 0.0 & 341,210 & 0.0\\
Less oil-supply disruption & 310,656 & $-$0.7 & 333,936 & $-$0.8 & 336,401 & $-$1.4\\
Less agg demand & 316,708 & 1.2 & 340,856 & 1.3 & 344,477 & 1.0\\
Less oil-specific demand & \textcolor{blue}{295,637} & \textcolor{blue}{$-$5.5} & \textcolor{blue}{314,912} & \textcolor{blue}{$-$6.4} & \textcolor{blue}{324,079} & \textcolor{blue}{$-$5.0}\\
Less local shocks & 298,324 & $-$4.6 & 324,276 & $-$3.6 & 349,211 & 2.3\\
\bottomrule
\end{tabular}
}

  \vspace{0.5em}
  
  \begin{tablenotesminipage}[Notes]
    Actual levels of employment are shown in the row ``\textbf{Employment data}.''
    The values are derived from the statistical model in \eqref{eq:svar} and constructed to
    match actual employment growth over the period January 2010 to December 2024.
    If the counterfactual series falls below actual employment,
    then the shock in question boosted employment.    
    The values in the table are shown in figure \ref{fig:historical-counterfactual-late}.
  \end{tablenotesminipage}
\end{table}

In contrast,
oil-supply and aggregate-demand shocks had a comparatively limited, yet consistent, influence on employment.
As seen in figure \ref{fig:historical-counterfactual-late} and reported in table \ref{tab:counterfactual-empl-late},
throughout the period,
unanticipated disruptions in crude-oil supply raised employment.
More often than not,
the dotted, green line associated with oil-supply shocks lies below the solid, pink line.
This pattern is visible in the row titled ``Less oil-supply disruption'' in table \ref{tab:counterfactual-empl-late}.
In January 2015
oil-supply shocks boosted employment by $0.7$ percent.
By January 2020, the boost amounted to $0.8$ percent and by December 2024 the boost amounted to 1.4 percent.

Aggregate-demand shocks consistently lowered employment.
In figure \ref{fig:historical-counterfactual-late},
the broken, blue line associated with aggregate-demand shocks lies above the solid, pink line associated with actual employment,
indicating employment would have been higher absent the influence of these shocks.
This pattern is reported in table \ref{tab:counterfactual-empl-late}.
In January 2015, January 2020, and December 2024,
aggregate-demand shocks pushed employment lower by roughly $1.2$, $1.3$, and $1$ percent.

Finally, 
regional-employment shocks meaningfully influenced employment.
This influence is visible in figure \ref{fig:historical-counterfactual-late}.
Until early $2020$,
shocks that were unattributable to the global crude-oil market boosted employment opportunities in Kern,
as the black, dashed line associated with local shocks lies below the solid, pink line associated with actual employment until early $2020$.
The influence of these shocks was greatest after $2015$ and before mid $2017$.
Table \ref{tab:counterfactual-empl-late} indicates that in January 2020 employment would have been $3.6$ percent lower absent the influence of these shocks.
Yet,
this pattern has recently reversed.
In December 2024,
the absence of these shocks would have boosted employment by $2.3$ percent,
indicating the local shock has become a drag on employment.

A main result of this exercise is demonstrating the importance of the oil market to Kern's economy.
Since $2010$,
the labor market in California's top oil-producing region has benefited from both unanticipated supply disruptions and oil-specific-demand shocks.
Kern may no longer benefit from these shocks as the US economy transitions away from fossil fuels.
This result indicates Kern's economy may be vulnerable to the green-energy transition.
On the other hand,
in the past, starting around $2015$,
Kern was able to add a significant number of jobs based on employment-demand shocks unrelated to the global market for crude oil. 
This result suggests resilience in Kern's economy.

\subsection{How much of the variation in Kern's employment growth can be attributed to each shock?}
\label{sec:fevd}

Using the estimated VAR model,
we can quantify how important the structural shocks are for employment growth. 
Table \ref{tab:fevd-dempl} 
reports how much of the forecast error variance is accounted for
at horizons of $1$, $2$, $3$, and $12$ months. 
Since employment growth is a stationary process,
a forecast horizon of $\infty$ indicates the variance decomposition of employment growth. 
Ignoring rounding error,
the entries in each row sum to $100$ percent.

\begin{table}

\caption{\label{tab:fevd-dempl} Forecast error variance decomposition for employment growth}
\centering
\begin{tabular}[t]{lllll}
\toprule
\multicolumn{1}{c}{ } & \multicolumn{4}{c}{Percent of $h$-step ahead forecast error variance explained by} \\
\cmidrule(l{3pt}r{3pt}){2-5}
\makecell[l]{\\Horizon} & \makecell[l]{Oil-supply\\shock} & \makecell[l]{Aggregate-\\demand shock} & \makecell[l]{Oil-specific-\\demand shock} & \makecell[l]{Residual\\shock}\\
\midrule
1 & 0.7 & 0.2 & 2.2 & 96.9\\
2 & 1.4 & 0.7 & 5.6 & 92.4\\
3 & 1.5 & 2.6 & 5.7 & 90.1\\
12 & 4.1 & 7.0 & 6.1 & 82.8\\
$\infty$ & 5.0 & 7.2 & 6.2 & 81.7\\
\bottomrule
\end{tabular}
\end{table}

Initially,
the oil-demand and -supply shocks combined explain little of the
mean squared prediction error of
employment growth.
Oil-specific-demand shocks explain $2.2$ percent of the variation in employment growth.
Oil-supply shocks explain $0.7$ percent.
Notably, supply shocks explain more of the variation than aggregate-demand shocks.

The explanatory power of oil-demand and -supply shocks rises to $7.2$ percent or more by $3$ months.
Around $80$ percent of this explanatory power is accounted for by oil-demand shocks.
Aggregate-demand shocks explain $2.6$ percent of the variation in employment growth.
The remaining variation is explained by Kern-specific shocks.
By one year, $10.2$ percent of the variation in employment growth is explained by oil-demand and -supply shocks.
Oil-supply shocks play an increasingly important role.

In the long run,
$7.2$ percent of the variance of employment growth is explained by aggregate-demand shocks.
These shocks account for more of the ups and downs of employment at longer horizons.
The oil market also explains a substantial fraction of the variance observed in employment growth:
$5$ percent of the variance of employment growth is explained by oil-supply shocks and
$6.2$ percent of the variance is explained by oil-demand shocks.
Overall, $11$ percent of the variance is explained by the oil market.

The result is surprising.
Even though Kern is California's top oil-producing region,
the industry directly employs few workers.
Since $1990$,
anywhere from $5.8$ to $1.5$ percent of workers
have been employed by firms engaged in oil-and-gas extraction or
firms that support oil-and-gas extraction.
The oil industry has an outsize effect on employment outcomes.\footnote{For completeness,
  we report forecast error variance decompositions for
percent changes in oil supply and the real price of oil in appendix \ref{sec:app:fevd}.}

\subsection{Understanding the effects of oil-price disturbances on wages}
\label{sec:wages}

Employment opportunities expand when the price of oil unexpectedly rises,
as indicated by the responses of employment in figure \ref{fig:irf-empl-block}.
A related question has to do with the quality of these opportunities.
One way to answer this question is looking at how average weekly wages respond to structural innovations
from the statistical model in \eqref{eq:svar}.
One challenge is data availability.
The QCEW reports average weekly wages at a quarterly frequency.
And while a comparable structural VAR model could be constructed using the wage data,
the identifying assumptions are not credible \citep{kilian_2009}.
For example, looking within the crude-oil block of the statistical model [the first three equations in the system shown in \eqref{eq:identification}],
a structural innovation to oil-specific demand may affect crude-oil supply \textit{within the quarter}.\footnote{The QCEW program publishes data on the average weekly wage.
  This statistic is the ratio of total compensation paid during the quarter to average employment over the quarter divided by 13, for the 13 weeks in the quarter.
  According to the BLS's \textit{Handbook of Methods},
  compensation includes
  bonuses, stock options, severance pay, profit distributions,
  the cash value of meals and lodging, tips, and other gratuities.
  In some states,
  employer contributions to certain deferred compensation plans are included.
  We deflate the series by the quarterly average of monthly CPI values.}

To overcome this challenge,
we follow \citet{kilian_2009} and average the structural shocks by quarter.
Letting $t$ now index quarter,
we define for the $j$th structural shock
\begin{align*}
\hat{\zeta}_{jt} = \frac{1}{3} \sum\limits_{\tau \in \text{ quarter }t}^{} \hat{\varepsilon}_{j\tau}.
\end{align*}
We investigate the relationship between real wage growth and the
quarterly structural innovations based on the regressions
\begin{equation}
\label{eq:second-stage-wage}
\Delta \log w_{t} = \alpha_{j} + \sum\limits_{i=0}^{8} \phi_{ji} \hat{\zeta}_{jt-i} + u_{jt},
\end{equation}
where $u_{jt}$ is a potentially serially correlated error.
The possibility for serial correlation in \eqref{eq:second-stage-wage} is accounted for
by constructing confidence intervals using a block-bootstrap procedure,
where the block length is chosen to be 4 quarters and we use $1,000$ bootstrap replications.\footnote{These confidence intervals do not account for
  the fact that the variables on the right side are generated regressors.}

We interpret
the regressions in \eqref{eq:second-stage-wage} under the assumption that,
within a given quarter,
structural innovations are predetermined with respect to real wage growth.
The term $\phi_{jh}$  in the regression model is the 
impulse response at horizon $h$ to structural innovation $j$.

Figure \ref{fig:wage-oil-supply-shock} shows the cumulative response of real wage growth
to oil-supply and oil-specific-demand shocks.
The top panel shows the cumulative response of real wage growth to
a negative oil-supply shock that will tend to raise the price of crude oil.
An unanticipated oil-supply disruption has little influence on real wages.
The bottom panel shows the cumulative response of real wage growth to
unanticipated oil-specific-demand increases.
Average real wages fall on impact and decline significantly by the first quarter.
They remain lower for around a year before returning to baseline.

\begin{figure}[htbp]
  \centering
  \resizebox{\textwidth}{!}{
    \input{./fig_plt_irf2_wages.tex}
  }
  \caption[]{\label{fig:wage-oil-supply-shock} Responses of the average real wage to
    oil-supply and oil-specific-demand shocks}
  \begin{figurenotes}[Notes]
    The plots show the cumulated responses estimated from the statistical model in \eqref{eq:second-stage-wage} along with
    one- and two-standard-error bands.
    Confidence intervals were constructed using a moving block bootstrap method described by \citet{kilian_2009}.
  \end{figurenotes}
\end{figure}

Recall that unanticipated disruptions to oil supply and unanticipated oil-specific-demand motives
expand employment opportunities, as shown in figure \ref{fig:irf-empl-block}.
Because it is reasonable to presume that expanded employment opportunities would raise wages of workers who do not exit employment,
the results in figure \ref{fig:wage-oil-supply-shock} suggest that workers added from nonemployment found jobs that paid below-average wages.
This also implies an unanticipated oil-specific-demand motive that lowers the price of crude oil
will raise the average wage as workers who earn lower wages exit employment.

The evidence that lower-wage workers who exit and enter employment in response to shocks in the global crude-oil market is a concern.
Much regional economic development and many place-based policies focus on ``good jobs'' \citep{miller-adams_etal_2019}.
As the fossil-fuel sector of Kern County, California, offers fewer opportunities, whether directly or indirectly,
the challenge will be connecting and preparing workers for new opportunities.
Kern resiliency may be tested if these opportunities are characterized by low pay.

Finally, 
the results, again, highlight the importance of differentiating
unanticipated supply disruptions from unanticipated precautionary-demand motives.

\section{Discussion}
\label{sec:discussion}

Parts of the United States are in the midst of an era of peak fossil-fuel extraction.
States like Texas and New Mexico are extracting more crude oil than ever before.
And states like Texas, Pennsylvania, and New Mexico are extracting more natural gas than ever before.
This revolution expanded employment opportunities.\footnote{See \citet{weber_2012} for early
  documentation. \citet{bartik_etal_2019} provide more recent evidence
  along with \citet{feyrer_mansur_sacerdote_2017} and
  the reinterpretation of \citeauthor{feyrer_mansur_sacerdote_2017}'s \citeyearpar{feyrer_mansur_sacerdote_2017} work by \citet{james_smith_2020} and \citet{feyrer_mansur_sacerdote_2020}.
  See also \citet{maniloff_mastromonaco_2017}.
  \citet{jacobsen_2019} and \citet{marchand_weber_2020} have investigated other aspects of fracking.
\citet{black_etal_2021} provide a general overview of this literature.}
Unlike these places, however, 
Kern did not experience a surge in fossil-fuel extraction associated with the fracking boom.

In many ways then,
Kern can be thought of as a pioneer in managing the transition with resilience.
From $2010$ to $2020$, for example,
factors unrelated to the global crude-oil market boosted Kern's employment meaningfully
(figure \ref{fig:historical-counterfactual-late} and table \ref{tab:counterfactual-empl-late}).
Nevertheless,
the transition away from fossil fuels
represents a significant reorganization of economic activity and regional factors
have recently been a drag on employment (the last column of table \ref{tab:counterfactual-empl-late}).
Episodes of past structural change portend struggle for
people in ``vulnerable'' regions like Kern
\citep{raimi_2021,raimi_carley_konisky_2022}.\footnote{In general,
  the consequences of structural change are not fully understood.
  There is a vast literature of these issues.
  We provide a brief overview here.
  \citet{rogerson_2015} offers a framework for evaluating the
  welfare effects of environmental regulation.
  \citet{davis_holladay_sims_2022} discuss the
  retirement of coal-fired power plants in the United States.
  \citet{hanson_2022,colmer_etal_2024} discuss
  the negative effects on workers of shutting down coal-fired power plants.
  \citet{acemoglu_autor_2011,acemoglu_restrepo_2020,acemoglu_restrepo_2022}
  investigate the automation of work.
  \citet{autor_dorn_hanson_2021,autor_dorn_hanson_2016}
  explore China's participation in world trade.
  The decline of middle-pay jobs and when this type of structural reallocation takes place over the business cycle is discussed by \citet{autor_dorn_2013,foote_ryan_2014,howes_2022}.
  \citet{herrendorf_rogerson_valentinyi_2014} and
  \citet{ngai_pissarides_2007}
  provide a framework 
  for structural change within the tradition of economic growth.}

Given how challenging it is to help regions that have previously
gone through structural change,
our findings
that link employment opportunities in Kern to the global oil market
take on more meaning.
Given that oil production is concentrated geographically,
there is added scope for place-based policies
\citep{austin_glaeser_summers_2018,bartik_2020jep,bartik_2020,kaufman_2024,raimi_witlock_2024}.

In addition to contributing to what is known about structural change,
our results also contribute to the growing literature on resilience.
\citet{modica_reggiani_2014} provide an early definition of this concept,
which was subsequently expanded by \citet{oesth_etal_2018}.
As \citet{de-cezaro-eberhardt_fochezatto_2024} emphasize,
one aim of this literature
is understanding why particular regions react differently to adverse shocks.
While their focus is on the recovery of Brazilian regions from the 2008
global financial crisis,
their points about heterogeneous recoveries apply here.
Many regions may be unaffected by the transition away from fossil fuels,
while others may be acutely exposed.
\citet{junchang_etal_2025}, for example,
highlight how oil revenue is key to understanding growth across Iran,
another region that depends on oil.

\section{Conclusion}
\label{sec:conclusion}

We estimated a structural VAR model that jointly explains the evolution of
key variables in
the global crude-oil market and the labor market in California's top
oil-producing region, Kern County.
Around $11$ percent of the variance in Kern's employment growth
can be attributed to the global crude-oil market.
Absent the influence of structural shocks associated with oil-specific demand since $2010$,
employment in Kern would be roughly $5$ percent lower.
These statistics suggest that there is scope for
the transition away from oil in California to be paired with place-based policies,
given that Kern accounts for $70$ percent of oil produced within the state.
These policies may well be effective.
Kern-specific influences contributed significantly to employment growth before
the Covid-19 recession.

Our statistical work was motivated by a theoretical framework of network-based production
that linked employment with public finances,
two main issues associated with the transition away from fossil fuels.
We used this theory to show how the impact of a change in the price of crude oil on sectoral employment can be
decomposed into three separate channels,
including a network-demand effect.
Input--output linkages, in general, will propagate and amplify shocks.

Kern may serve as a test case for other regions.
Unlike California,
Texas and New Mexico are in the midst of peak oil production.
How Kern manages the transition 
may well serve as a test case for other regions.
Yet,
designing a transition that maximizes flourishing will require
being open minded about how to help communities---a one-size-fits-all policy seems inadequate.
Understanding how local economies have benefited from oil and local shocks,
like we did in this paper,
may be a useful tool for policymakers.

\appendix
\section{Data on Kern County, California: employment in fossil-fuel extraction, per-capita personal income, educational attainment}
\label{sec:appendix:data-kern}

In this section,
we share some background data on Kern County, California.
Data on employment in fossil-fuel extraction are presented in section \ref{sec:empl-foss-fuel}.
Data on per-capita personal income across California counties in $2024$ are presented in section \ref{sec:pcpi}.
Data on educational attainment across California counties are presented in section \ref{sec:educ-atta}.

\subsection{Employment in fossil-fuel extraction}
\label{sec:empl-foss-fuel}

\begin{figure}[htbp]
  \centering
  \resizebox{\textwidth}{!}{
  \input{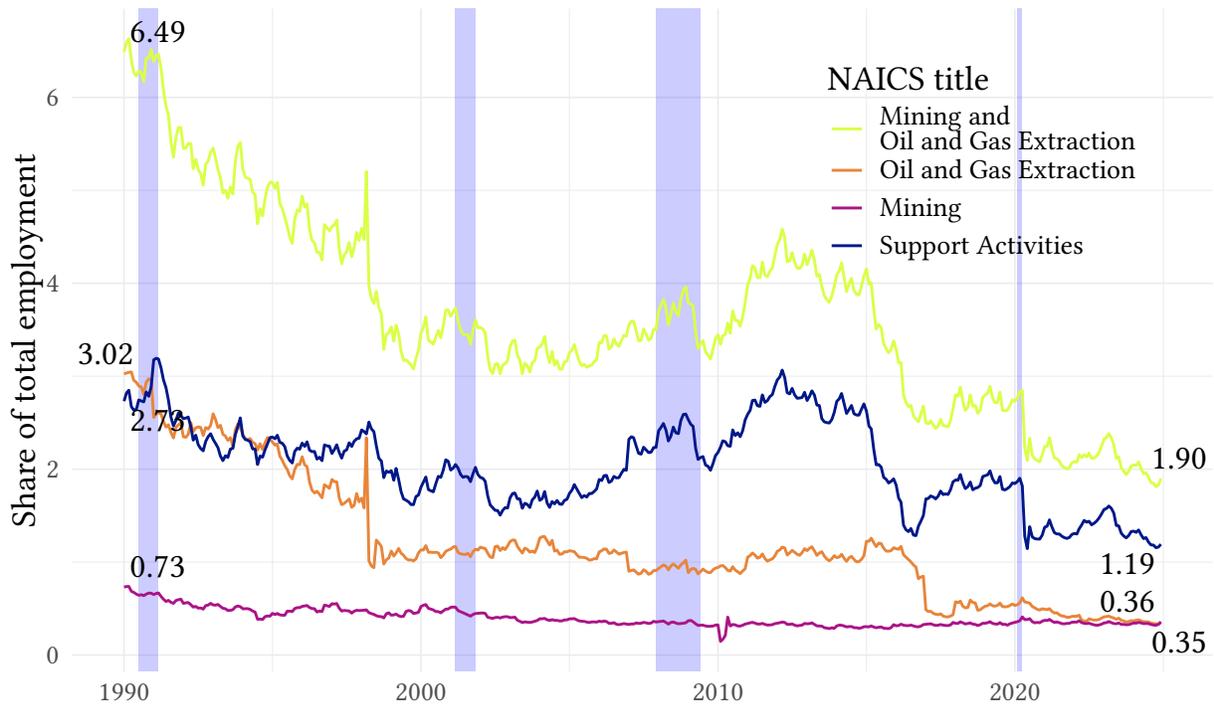}}
\caption{Share of employment in select industries, Kern County, California, January 1990 through December 2024}
\label{fig:empl-share-oil}
\begin{figurenotes}[Notes]
  The mining and oil-and-gas extraction industry (shown in yellow)
  comprises three subindustries:
  oil and gas extraction (shown in orange),
  mining (except oil and gas extraction, shown in maroon), and
  support activities for these two sectors (shown in blue).
\end{figurenotes}
\begin{figurenotes}[Source]
  Quarterly Census of Employment and Wages.
\end{figurenotes}
\end{figure}

In Kern County, California,
the fossil-fuel sector directly employs few workers.
This can be seen in 
data from the Quarterly Census of Employment and Wages.
Employment data developed by the QCEW program classify workers into detailed industries.
The industrial classification includes the two-digit industry code mining, quarrying, and oil and gas extraction.
The share of Kern's employment in this industry is shown in figure \ref{fig:empl-share-oil}.

At the start of $1990$,
less than $6.5$ percent of jobs were in this sector. 
By the end of $2024$,
less than $2$ percent of jobs were in this sector.
These statistics represent an upper bound on the share of workers directly engaged in fossil-fuel extraction.
The two-digit industry comprises
workers engaged in oil-and-gas extraction, mining, and support activities for these two industries.
These series are also shown in figure \ref{fig:empl-share-oil}.

\subsection{Per-capita personal income across California counties in 2024}
\label{sec:pcpi}

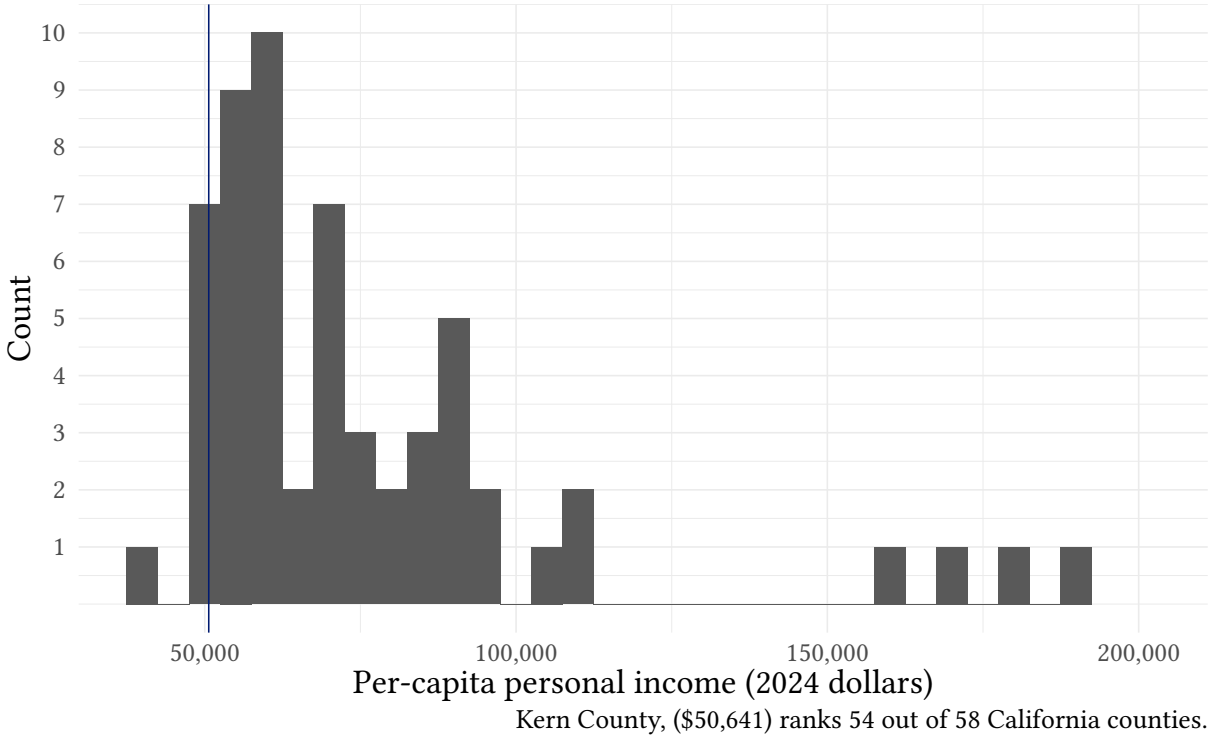
\begin{figure}[htbp]
  \centering
  \resizebox{\textwidth}{!}{
\begin{tikzpicture}[x=1pt,y=1pt]
\definecolor{fillColor}{RGB}{255,255,255}
\path[use as bounding box,fill=fillColor,fill opacity=0.00] (0,0) rectangle (469.75,290.32);
\begin{scope}
\path[clip] ( 31.71, 43.96) rectangle (464.25,284.82);
\definecolor{drawColor}{gray}{0.92}

\path[draw=drawColor,line width= 0.3pt,line join=round] ( 31.71, 54.91) --
	(464.25, 54.91);

\path[draw=drawColor,line width= 0.3pt,line join=round] ( 31.71, 65.85) --
	(464.25, 65.85);

\path[draw=drawColor,line width= 0.3pt,line join=round] ( 31.71, 87.75) --
	(464.25, 87.75);

\path[draw=drawColor,line width= 0.3pt,line join=round] ( 31.71,109.65) --
	(464.25,109.65);

\path[draw=drawColor,line width= 0.3pt,line join=round] ( 31.71,131.55) --
	(464.25,131.55);

\path[draw=drawColor,line width= 0.3pt,line join=round] ( 31.71,153.44) --
	(464.25,153.44);

\path[draw=drawColor,line width= 0.3pt,line join=round] ( 31.71,175.34) --
	(464.25,175.34);

\path[draw=drawColor,line width= 0.3pt,line join=round] ( 31.71,197.24) --
	(464.25,197.24);

\path[draw=drawColor,line width= 0.3pt,line join=round] ( 31.71,219.13) --
	(464.25,219.13);

\path[draw=drawColor,line width= 0.3pt,line join=round] ( 31.71,241.03) --
	(464.25,241.03);

\path[draw=drawColor,line width= 0.3pt,line join=round] ( 31.71,262.93) --
	(464.25,262.93);

\path[draw=drawColor,line width= 0.3pt,line join=round] ( 31.71,284.82) --
	(464.25,284.82);

\path[draw=drawColor,line width= 0.3pt,line join=round] (139.64, 43.96) --
	(139.64,284.82);

\path[draw=drawColor,line width= 0.3pt,line join=round] (258.90, 43.96) --
	(258.90,284.82);

\path[draw=drawColor,line width= 0.3pt,line join=round] (378.15, 43.96) --
	(378.15,284.82);

\path[draw=drawColor,line width= 0.6pt,line join=round] ( 31.71, 76.80) --
	(464.25, 76.80);

\path[draw=drawColor,line width= 0.6pt,line join=round] ( 31.71, 98.70) --
	(464.25, 98.70);

\path[draw=drawColor,line width= 0.6pt,line join=round] ( 31.71,120.60) --
	(464.25,120.60);

\path[draw=drawColor,line width= 0.6pt,line join=round] ( 31.71,142.49) --
	(464.25,142.49);

\path[draw=drawColor,line width= 0.6pt,line join=round] ( 31.71,164.39) --
	(464.25,164.39);

\path[draw=drawColor,line width= 0.6pt,line join=round] ( 31.71,186.29) --
	(464.25,186.29);

\path[draw=drawColor,line width= 0.6pt,line join=round] ( 31.71,208.18) --
	(464.25,208.18);

\path[draw=drawColor,line width= 0.6pt,line join=round] ( 31.71,230.08) --
	(464.25,230.08);

\path[draw=drawColor,line width= 0.6pt,line join=round] ( 31.71,251.98) --
	(464.25,251.98);

\path[draw=drawColor,line width= 0.6pt,line join=round] ( 31.71,273.88) --
	(464.25,273.88);

\path[draw=drawColor,line width= 0.6pt,line join=round] ( 80.01, 43.96) --
	( 80.01,284.82);

\path[draw=drawColor,line width= 0.6pt,line join=round] (199.27, 43.96) --
	(199.27,284.82);

\path[draw=drawColor,line width= 0.6pt,line join=round] (318.52, 43.96) --
	(318.52,284.82);

\path[draw=drawColor,line width= 0.6pt,line join=round] (437.78, 43.96) --
	(437.78,284.82);
\definecolor{fillColor}{gray}{0.35}

\path[fill=fillColor] ( 50.20, 54.91) rectangle ( 62.12, 76.80);

\path[fill=fillColor] ( 62.12, 54.91) rectangle ( 74.05, 54.91);

\path[fill=fillColor] ( 74.05, 54.91) rectangle ( 85.97,208.18);

\path[fill=fillColor] ( 85.97, 54.91) rectangle ( 97.90,251.98);

\path[fill=fillColor] ( 97.90, 54.91) rectangle (109.83,273.88);

\path[fill=fillColor] (109.83, 54.91) rectangle (121.75, 98.70);

\path[fill=fillColor] (121.75, 54.91) rectangle (133.68,208.18);

\path[fill=fillColor] (133.68, 54.91) rectangle (145.60,120.60);

\path[fill=fillColor] (145.60, 54.91) rectangle (157.53, 98.70);

\path[fill=fillColor] (157.53, 54.91) rectangle (169.45,120.60);

\path[fill=fillColor] (169.45, 54.91) rectangle (181.38,164.39);

\path[fill=fillColor] (181.38, 54.91) rectangle (193.30, 98.70);

\path[fill=fillColor] (193.30, 54.91) rectangle (205.23, 54.91);

\path[fill=fillColor] (205.23, 54.91) rectangle (217.16, 76.80);

\path[fill=fillColor] (217.16, 54.91) rectangle (229.08, 98.70);

\path[fill=fillColor] (229.08, 54.91) rectangle (241.01, 54.91);

\path[fill=fillColor] (241.01, 54.91) rectangle (252.93, 54.91);

\path[fill=fillColor] (252.93, 54.91) rectangle (264.86, 54.91);

\path[fill=fillColor] (264.86, 54.91) rectangle (276.78, 54.91);

\path[fill=fillColor] (276.78, 54.91) rectangle (288.71, 54.91);

\path[fill=fillColor] (288.71, 54.91) rectangle (300.64, 54.91);

\path[fill=fillColor] (300.64, 54.91) rectangle (312.56, 54.91);

\path[fill=fillColor] (312.56, 54.91) rectangle (324.49, 54.91);

\path[fill=fillColor] (324.49, 54.91) rectangle (336.41, 54.91);

\path[fill=fillColor] (336.41, 54.91) rectangle (348.34, 76.80);

\path[fill=fillColor] (348.34, 54.91) rectangle (360.26, 54.91);

\path[fill=fillColor] (360.26, 54.91) rectangle (372.19, 76.80);

\path[fill=fillColor] (372.19, 54.91) rectangle (384.11, 54.91);

\path[fill=fillColor] (384.11, 54.91) rectangle (396.04, 76.80);

\path[fill=fillColor] (396.04, 54.91) rectangle (407.97, 54.91);

\path[fill=fillColor] (407.97, 54.91) rectangle (419.89, 76.80);
\definecolor{drawColor}{RGB}{0,26,112}

\path[draw=drawColor,line width= 0.6pt,line join=round] ( 81.54, 43.96) -- ( 81.54,284.82);
\end{scope}
\begin{scope}
\path[clip] (  0.00,  0.00) rectangle (469.75,290.32);
\definecolor{drawColor}{gray}{0.30}

\node[text=drawColor,anchor=base east,inner sep=0pt, outer sep=0pt, scale=  0.88] at ( 26.76, 73.77) {1};

\node[text=drawColor,anchor=base east,inner sep=0pt, outer sep=0pt, scale=  0.88] at ( 26.76, 95.67) {2};

\node[text=drawColor,anchor=base east,inner sep=0pt, outer sep=0pt, scale=  0.88] at ( 26.76,117.57) {3};

\node[text=drawColor,anchor=base east,inner sep=0pt, outer sep=0pt, scale=  0.88] at ( 26.76,139.46) {4};

\node[text=drawColor,anchor=base east,inner sep=0pt, outer sep=0pt, scale=  0.88] at ( 26.76,161.36) {5};

\node[text=drawColor,anchor=base east,inner sep=0pt, outer sep=0pt, scale=  0.88] at ( 26.76,183.26) {6};

\node[text=drawColor,anchor=base east,inner sep=0pt, outer sep=0pt, scale=  0.88] at ( 26.76,205.15) {7};

\node[text=drawColor,anchor=base east,inner sep=0pt, outer sep=0pt, scale=  0.88] at ( 26.76,227.05) {8};

\node[text=drawColor,anchor=base east,inner sep=0pt, outer sep=0pt, scale=  0.88] at ( 26.76,248.95) {9};

\node[text=drawColor,anchor=base east,inner sep=0pt, outer sep=0pt, scale=  0.88] at ( 26.76,270.85) {10};
\end{scope}
\begin{scope}
\path[clip] (  0.00,  0.00) rectangle (469.75,290.32);
\definecolor{drawColor}{gray}{0.30}

\node[text=drawColor,anchor=base,inner sep=0pt, outer sep=0pt, scale=  0.88] at ( 80.01, 32.95) {50,000};

\node[text=drawColor,anchor=base,inner sep=0pt, outer sep=0pt, scale=  0.88] at (199.27, 32.95) {100,000};

\node[text=drawColor,anchor=base,inner sep=0pt, outer sep=0pt, scale=  0.88] at (318.52, 32.95) {150,000};

\node[text=drawColor,anchor=base,inner sep=0pt, outer sep=0pt, scale=  0.88] at (437.78, 32.95) {200,000};
\end{scope}
\begin{scope}
\path[clip] (  0.00,  0.00) rectangle (469.75,290.32);
\definecolor{drawColor}{RGB}{0,0,0}

\node[text=drawColor,anchor=base,inner sep=0pt, outer sep=0pt, scale=  1.10] at (247.98, 20.91) {Per-capita personal income (2024 dollars)};
\end{scope}
\begin{scope}
\path[clip] (  0.00,  0.00) rectangle (469.75,290.32);
\definecolor{drawColor}{RGB}{0,0,0}

\node[text=drawColor,rotate= 90.00,anchor=base,inner sep=0pt, outer sep=0pt, scale=  1.10] at ( 13.08,164.39) {Count};
\end{scope}
\begin{scope}
\path[clip] (  0.00,  0.00) rectangle (469.75,290.32);
\definecolor{drawColor}{RGB}{0,0,0}

\node[text=drawColor,anchor=base east,inner sep=0pt, outer sep=0pt, scale=  0.88] at (464.25,  7.21) {Kern County (\$50,641) ranks 54 out of 58 California counties.};
\end{scope}
\end{tikzpicture}}  
  \caption{Summary of per-capita personal income across California counties}
  \label{fig:pcpi-ca}
  \begin{figurenotes}[Source]
U.S.~Bureau of Economic Analysis,
"\href{https://apps.bea.gov/itable/?ReqID=70&step=1&_gl=1*1lvyi86*_ga*Mjg5MTEwNjgyLjE3NzE3ODIyMjg.*_ga_J4698JNNFT*czE3NzE3ODIyMjgkbzEkZzEkdDE3NzE3ODI0ODEkajEwJGwwJGgw#eyJhcHBpZCI6NzAsInN0ZXBzIjpbMSwyOSwyNSwzMSwyNiwyNywzMF0sImRhdGEiOltbIlRhYmxlSWQiLCIyMCJdLFsiTWFqb3JfQXJlYSIsIjQiXSxbIlN0YXRlIixbIlhYIl1dLFsiQXJlYSIsWyJYWCJdXSxbIlN0YXRpc3RpYyIsWyIzIl1dLFsiVW5pdF9vZl9tZWFzdXJlIiwiTGV2ZWxzIl0sWyJZZWFyIixbIjIwMjQiXV0sWyJZZWFyQmVnaW4iLCItMSJdLFsiWWVhcl9FbmQiLCItMSJdXX0=}{CAINC1 County personal income summary: personal income, population, per capita personal income}." Accessed on February 22, 2026.    
  \end{figurenotes}
\end{figure}

The Bureau of Economic Analysis reports county-level data as part of its regional economic accounts,
including data on personal income.
Personal income
consists of income received in return for the provision of labor, land, and capital in addition to transfer receipts. 
Figure \ref{fig:pcpi-ca} reports the histogram of personal income divided by resident population for $58$ California counties.
All the data are from $2024$.
If the $58$ counties in California are ordered by per-capita personal income, from highest to lowest,
Kern would rank $54$.
Kern's level of per-capita personal income is shown in figure \ref{fig:pcpi-ca} using a blue, vertical line.

\subsection{Educational attainment across California counties}
\label{sec:educ-atta}

\begin{figure}[htbp]
  \centering
  \resizebox{\textwidth}{!}{
\begin{tikzpicture}[x=1pt,y=1pt]
\definecolor{fillColor}{RGB}{255,255,255}
\path[use as bounding box,fill=fillColor,fill opacity=0.00] (0,0) rectangle (469.75,290.32);
\begin{scope}
\path[clip] ( 31.71, 43.96) rectangle (464.25,284.82);
\definecolor{drawColor}{gray}{0.92}

\path[draw=drawColor,line width= 0.3pt,line join=round] ( 31.71, 54.91) --
	(464.25, 54.91);

\path[draw=drawColor,line width= 0.3pt,line join=round] ( 31.71, 65.85) --
	(464.25, 65.85);

\path[draw=drawColor,line width= 0.3pt,line join=round] ( 31.71, 87.75) --
	(464.25, 87.75);

\path[draw=drawColor,line width= 0.3pt,line join=round] ( 31.71,109.65) --
	(464.25,109.65);

\path[draw=drawColor,line width= 0.3pt,line join=round] ( 31.71,131.55) --
	(464.25,131.55);

\path[draw=drawColor,line width= 0.3pt,line join=round] ( 31.71,153.44) --
	(464.25,153.44);

\path[draw=drawColor,line width= 0.3pt,line join=round] ( 31.71,175.34) --
	(464.25,175.34);

\path[draw=drawColor,line width= 0.3pt,line join=round] ( 31.71,197.24) --
	(464.25,197.24);

\path[draw=drawColor,line width= 0.3pt,line join=round] ( 31.71,219.13) --
	(464.25,219.13);

\path[draw=drawColor,line width= 0.3pt,line join=round] ( 31.71,241.03) --
	(464.25,241.03);

\path[draw=drawColor,line width= 0.3pt,line join=round] ( 31.71,262.93) --
	(464.25,262.93);

\path[draw=drawColor,line width= 0.3pt,line join=round] ( 31.71,284.82) --
	(464.25,284.82);

\path[draw=drawColor,line width= 0.3pt,line join=round] ( 42.09, 43.96) --
	( 42.09,284.82);

\path[draw=drawColor,line width= 0.3pt,line join=round] (195.56, 43.96) --
	(195.56,284.82);

\path[draw=drawColor,line width= 0.3pt,line join=round] (349.03, 43.96) --
	(349.03,284.82);

\path[draw=drawColor,line width= 0.6pt,line join=round] ( 31.71, 76.80) --
	(464.25, 76.80);

\path[draw=drawColor,line width= 0.6pt,line join=round] ( 31.71, 98.70) --
	(464.25, 98.70);

\path[draw=drawColor,line width= 0.6pt,line join=round] ( 31.71,120.60) --
	(464.25,120.60);

\path[draw=drawColor,line width= 0.6pt,line join=round] ( 31.71,142.49) --
	(464.25,142.49);

\path[draw=drawColor,line width= 0.6pt,line join=round] ( 31.71,164.39) --
	(464.25,164.39);

\path[draw=drawColor,line width= 0.6pt,line join=round] ( 31.71,186.29) --
	(464.25,186.29);

\path[draw=drawColor,line width= 0.6pt,line join=round] ( 31.71,208.18) --
	(464.25,208.18);

\path[draw=drawColor,line width= 0.6pt,line join=round] ( 31.71,230.08) --
	(464.25,230.08);

\path[draw=drawColor,line width= 0.6pt,line join=round] ( 31.71,251.98) --
	(464.25,251.98);

\path[draw=drawColor,line width= 0.6pt,line join=round] ( 31.71,273.88) --
	(464.25,273.88);

\path[draw=drawColor,line width= 0.6pt,line join=round] (118.83, 43.96) --
	(118.83,284.82);

\path[draw=drawColor,line width= 0.6pt,line join=round] (272.30, 43.96) --
	(272.30,284.82);

\path[draw=drawColor,line width= 0.6pt,line join=round] (425.77, 43.96) --
	(425.77,284.82);
\definecolor{fillColor}{gray}{0.35}

\path[fill=fillColor] ( 51.37, 54.91) rectangle ( 75.95, 98.70);

\path[fill=fillColor] ( 75.95, 54.91) rectangle (100.53,164.39);

\path[fill=fillColor] (100.53, 54.91) rectangle (125.10,186.29);

\path[fill=fillColor] (125.10, 54.91) rectangle (149.68,273.88);

\path[fill=fillColor] (149.68, 54.91) rectangle (174.25,208.18);

\path[fill=fillColor] (174.25, 54.91) rectangle (198.83, 98.70);

\path[fill=fillColor] (198.83, 54.91) rectangle (223.41,164.39);

\path[fill=fillColor] (223.41, 54.91) rectangle (247.98,142.49);

\path[fill=fillColor] (247.98, 54.91) rectangle (272.56,142.49);

\path[fill=fillColor] (272.56, 54.91) rectangle (297.14,142.49);

\path[fill=fillColor] (297.14, 54.91) rectangle (321.71,142.49);

\path[fill=fillColor] (321.71, 54.91) rectangle (346.29, 54.91);

\path[fill=fillColor] (346.29, 54.91) rectangle (370.87, 76.80);

\path[fill=fillColor] (370.87, 54.91) rectangle (395.44, 76.80);

\path[fill=fillColor] (395.44, 54.91) rectangle (420.02, 76.80);

\path[fill=fillColor] (420.02, 54.91) rectangle (444.59, 98.70);
\definecolor{drawColor}{RGB}{0,26,112}

\path[draw=drawColor,line width= 0.6pt,line join=round] (109.40, 43.96) -- (109.40,284.82);
\end{scope}
\begin{scope}
\path[clip] (  0.00,  0.00) rectangle (469.75,290.32);
\definecolor{drawColor}{gray}{0.30}

\node[text=drawColor,anchor=base east,inner sep=0pt, outer sep=0pt, scale=  0.88] at ( 26.76, 73.77) {1};

\node[text=drawColor,anchor=base east,inner sep=0pt, outer sep=0pt, scale=  0.88] at ( 26.76, 95.67) {2};

\node[text=drawColor,anchor=base east,inner sep=0pt, outer sep=0pt, scale=  0.88] at ( 26.76,117.57) {3};

\node[text=drawColor,anchor=base east,inner sep=0pt, outer sep=0pt, scale=  0.88] at ( 26.76,139.46) {4};

\node[text=drawColor,anchor=base east,inner sep=0pt, outer sep=0pt, scale=  0.88] at ( 26.76,161.36) {5};

\node[text=drawColor,anchor=base east,inner sep=0pt, outer sep=0pt, scale=  0.88] at ( 26.76,183.26) {6};

\node[text=drawColor,anchor=base east,inner sep=0pt, outer sep=0pt, scale=  0.88] at ( 26.76,205.15) {7};

\node[text=drawColor,anchor=base east,inner sep=0pt, outer sep=0pt, scale=  0.88] at ( 26.76,227.05) {8};

\node[text=drawColor,anchor=base east,inner sep=0pt, outer sep=0pt, scale=  0.88] at ( 26.76,248.95) {9};

\node[text=drawColor,anchor=base east,inner sep=0pt, outer sep=0pt, scale=  0.88] at ( 26.76,270.85) {10};
\end{scope}
\begin{scope}
\path[clip] (  0.00,  0.00) rectangle (469.75,290.32);
\definecolor{drawColor}{gray}{0.30}

\node[text=drawColor,anchor=base,inner sep=0pt, outer sep=0pt, scale=  0.88] at (118.83, 32.95) {0.2};

\node[text=drawColor,anchor=base,inner sep=0pt, outer sep=0pt, scale=  0.88] at (272.30, 32.95) {0.4};

\node[text=drawColor,anchor=base,inner sep=0pt, outer sep=0pt, scale=  0.88] at (425.77, 32.95) {0.6};
\end{scope}
\begin{scope}
\path[clip] (  0.00,  0.00) rectangle (469.75,290.32);
\definecolor{drawColor}{RGB}{0,0,0}

\node[text=drawColor,anchor=base,inner sep=0pt, outer sep=0pt, scale=  1.10] at (247.98, 20.91) {Share of people who have earned a college degree};
\end{scope}
\begin{scope}
\path[clip] (  0.00,  0.00) rectangle (469.75,290.32);
\definecolor{drawColor}{RGB}{0,0,0}

\node[text=drawColor,rotate= 90.00,anchor=base,inner sep=0pt, outer sep=0pt, scale=  1.10] at ( 13.08,164.39) {Count};
\end{scope}
\begin{scope}
\path[clip] (  0.00,  0.00) rectangle (469.75,290.32);
\definecolor{drawColor}{RGB}{0,0,0}

\node[text=drawColor,anchor=base east,inner sep=0pt, outer sep=0pt, scale=  0.88] at (464.25,  7.21) {Kern County (19 percent) ranks 48 out of 58 California counties.};
\end{scope}
\end{tikzpicture}}  
  \caption{Summary of educational attainment across California counties}
  \label{fig:edu-ca}
  \begin{figurenotes}[Source]
    Authors' calculations using data from the
U.S.~Census Bureau. "Sex by Educational Attainment for the Population 25 Years and Over." American Community Survey, ACS 5-Year Estimates Detailed Tables, 
\href{https://data.census.gov/table/ACSDT5Y2024.B15002?t=Educational+Attainment&g=040XX00US06$0500000_050XX00US24001,24003&moe=false&tp=false&tableFilters=ag-Grid-AutoColumn~(Margin+of+Errorundefined)}{Table B15002}.
Accessed on February 22, 2026.    
  \end{figurenotes}
\end{figure}
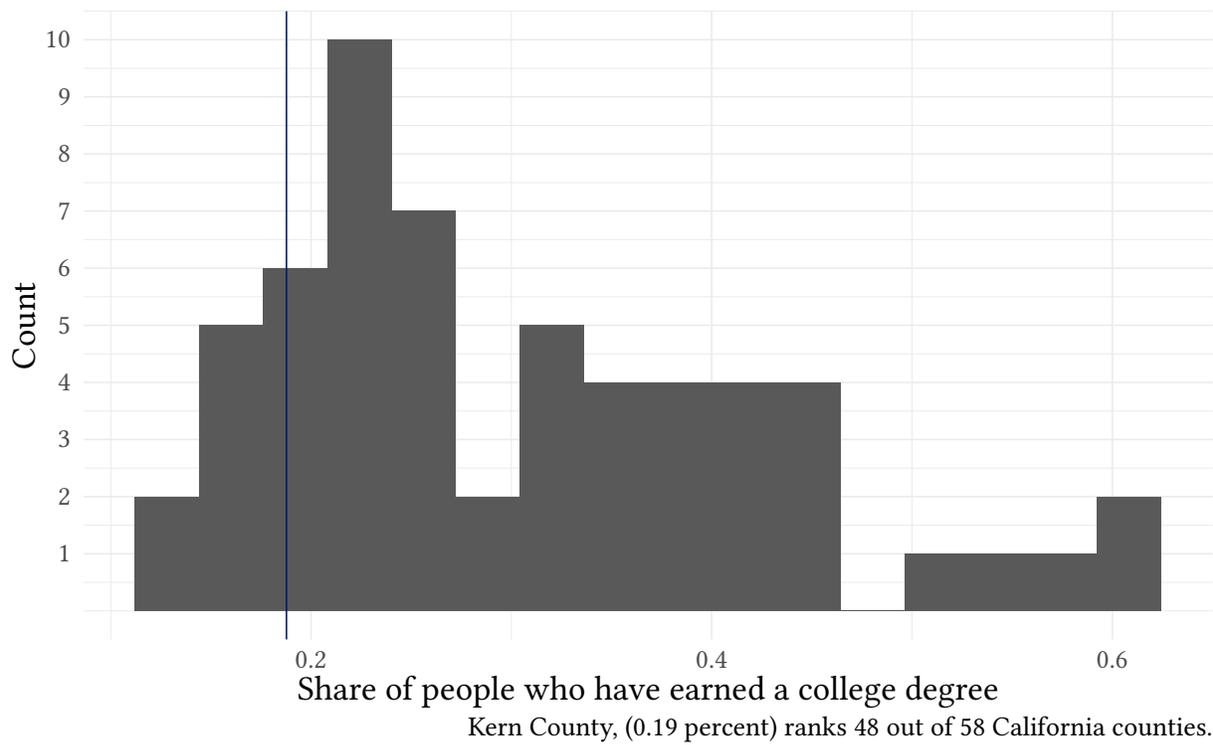

Residents of Kern County, California, have earned significantly less education than other residents of California.
Using data from the American Community Survey,
we report the share of
people aged 25 years and over who have earned a Bachelor's degree, a Master's degree, professional degree, or a doctoral degree.
We do this by California county in figure \ref{fig:edu-ca} using data from $2024$.
Less than $20$ percent of people in Kern County have earned at least a college degree.
Across the $58$ counties this ranks $48$th.
Nationally, closer to $39$ percent of people aged at least $25$ have earned at least a college degree.

\section{Forecast error variance decompositions}
\label{sec:app:fevd}

For completeness,
we also report forecast error variance decompositions for
crude-oil production and the real price of oil.
The forecast error variance decompositions answer
the question of how much of the variability of
the percent change of global crude-oil production and
the log of the real price of crude oil can be explained
by oil-supply, oil-demand, and residual shocks.
Since $\Delta \; \text{oil production}$ and the real price of oil
are stationary variables, 
we consider a forecast horizon of $h=\infty$,
which is the variance decomposition,
along with horizons of length $1$, $2$, $3$, and $12$.
The decompositions are shown in tables \ref{tab:fevd-oilsupply} and \ref{tab:fevd-rpoil}.

As seen in table \ref{tab:fevd-oilsupply},
at the one-month horizon, $100$ percent of the forecast error variance
in crude-oil production is explained by the oil-supply shock.
This reflects the recursive identification scheme in \eqref{eq:identification}.
At longer horizons,
the explanatory power of oil-supply shocks decreases;
yet, roughly two thirds of the variance in the long run is explained by unanticipated oil-supply disruption.
Demand shocks have explanatory power after a delay.
By one year,
aggregate-demand shocks explain $6.4$ percent and oil-specific-demand shocks explain $7.8$ percent of the mean square prediction error.
These shares of explanatory power remain similar in the long run.
The remaining portion of the long-run variance is
explained by forces outside of the global crude-oil market.
Overall,
table \ref{tab:fevd-oilsupply} supports \citeauthor{kilian_2009}'s \citeyearpar{kilian_2009} interpretation that demand conditions influence changes in crude-oil supply only after some adjustment period.

\begin{table}

\caption{\label{tab:fevd-oilsupply} Forecast error variance decomposition for oil supply}
\centering
\begin{tabular}[t]{lllll}
\toprule
\multicolumn{1}{c}{ } & \multicolumn{4}{c}{Percent of $h$-step ahead forecast error variance explained by} \\
\cmidrule(l{3pt}r{3pt}){2-5}
\makecell[l]{\\Horizon} & \makecell[l]{Oil-supply\\shock} & \makecell[l]{Aggregate-\\demand shock} & \makecell[l]{Oil-specific-\\demand shock} & \makecell[l]{Residual\\shock}\\
\midrule
1 & 100.0 & 0.0 & 0.0 & 0.0\\
2 & 81.1 & 0.1 & 1.9 & 16.8\\
3 & 76.8 & 0.3 & 7.2 & 15.7\\
12 & 69.2 & 6.4 & 7.8 & 16.6\\
$\infty$ & 68.4 & 7.0 & 8.1 & 16.6\\
\bottomrule
\end{tabular}
\end{table}

The forecast error variance decomposition of the real price of crude oil
(reported in table \ref{tab:fevd-rpoil})
indicates that variation in the real price of crude oil is overwhelmingly explained by
factors other than supply, a main message of \citet{kilian_2009}.
On impact,
nearly all the mean square prediction error is explained by oil-specific-demand shocks.
This explanatory power falls over time.
In the long run,
around half of the variation of the real price of oil is explained by oil-specific-demand shocks.
Whereas,
aggregate-demand and oil-supply shocks explain $26.2$ and $9.7$ percent of the variability of the real price of crude oil.
Finally, only $8$ percent of the variability of the real price of oil is explained by factors outside of the global crude-oil market.

\begin{table}

\caption{\label{tab:fevd-rpoil} Forecast error variance decomposition for the real price of crude oil}
\centering
\begin{tabular}[t]{lllll}
\toprule
\multicolumn{1}{c}{ } & \multicolumn{4}{c}{Percent of $h$-step ahead forecast error variance explained by} \\
\cmidrule(l{3pt}r{3pt}){2-5}
\makecell[l]{\\Horizon} & \makecell[l]{Oil-supply\\shock} & \makecell[l]{Aggregate-\\demand shock} & \makecell[l]{Oil-specific-\\demand shock} & \makecell[l]{Residual\\shock}\\
\midrule
1 & 1.7 & 1.3 & 97.1 & 0.0\\
2 & 1.3 & 3.4 & 94.6 & 0.7\\
3 & 1.1 & 6.6 & 90.0 & 2.3\\
12 & 5.3 & 14.2 & 76.1 & 4.4\\
$\infty$ & 9.7 & 26.2 & 56.0 & 8.0\\
\bottomrule
\end{tabular}
\end{table}

Together,
tables \ref{tab:fevd-oilsupply} and \ref{tab:fevd-rpoil}
describe a key feature of the estimated crude-oil block:
supply fluctuations are mainly explained by supply shocks; whereas,
price fluctuations are mainly explained by demand shocks.
This evidence is consistent with insights made by \citet{kilian_2009}.

\section{Derivation of sectoral-employment changes in response to changes in the price oil}
\label{sec:app:model}

In this appendix,
we provide further details about the theoretical model in section \ref{sec:theory}.

\subsection{Firms' choices about profit maximization and cost minimization}
\label{sec:firms-choices-about-profit-max-cost-min}

A representative firm in sector $j$ maximizes profits, taking prices as given:
\begin{align*}
\max_{l_{j},\left\{ x_{ji}\right\} _{i=1}^{n}}\quad p_{j}l_{j}^{\alpha_{j}^{l}}\prod_{i=1}^{n}x_{ji}^{a_{ji}}-\sum_{i=1}^{n}p_{i}x_{ji}-wl_{j}.
\end{align*}
The firm's choice of input $x_{ij}$ is summarized by the first-order condition and implies
\begin{equation}
\label{eq:profit-max-choice-xji}
x_{ji} =\frac{a_{ji}p_{j}y_{j}}{p_{i}}.
\end{equation}
The firm's choice of labor is summarized by the first-order condition and implies
\begin{equation}
  \label{eq:profit-max-choice-labor}
l_{j} =\frac{\alpha_{j}^{l}p_{j}y_{j}}{w}.
\end{equation}
A representative firm's choices about inputs expressed in \eqref{eq:profit-max-choice-xji} indicates
\begin{equation}
a_{ij}=\frac{p_{j}x_{ij}}{p_{i}y_{i}},\label{eq:def-aij}
\end{equation}
which is the input expenditure on input $j$, $p_{j}x_{ij}$, as a fraction of total sales, $p_{i}y_{i}$.
These are components of the regional economy's input--output matrix $\bm{A}=\left[a_{ij}\right]$. 

\subsection{Unit cost function}

For a representative firm in sector $j$,
the Lagrangian associated with the unit-cost problem is
\begin{align*}
L_{j}^{\text{unit cost}}=\sum_{k=1}^{n}p_{k}x_{jk}+wl_{j}+\lambda^{f}\left(1-l_{j}^{\alpha_{i}^{l}}\prod_{k=1}^{n}x_{jk}^{a_{jk}}\right).
\end{align*}
Using the binding constraint that output is $1$,
the first-order condition for $x_{jk}$ can be expressed as
\begin{equation}
p_{k}x_{jk}=\lambda^{f}a_{jk}.\label{eq:min-cost-foc-x}
\end{equation}
The first-order condition for $l_{j}$, using the binding constraint that output is $1$, can be expressed as
\begin{equation}
wl_{j}=\lambda^{f}\alpha_{j}^{l}.\label{eq:min-cost-foc-l}
\end{equation}
The two optimality conditions in \eqref{eq:min-cost-foc-x} and \eqref{eq:min-cost-foc-l} imply an expression for $\lambda^{f}$ in terms of unit cost; indeed,
\begin{align*}
wl_{j}+\sum_{k=1}^{n}p_{k}x_{jk}=\lambda^{f}\left(\alpha_{j}^{l}+\sum_{k=1}^{n}a_{jk}\right)=\lambda^{f},
\end{align*}
where the last equality uses the constant-returns-to-scale assumption in \eqref{eq:technology-params}; namely,
$\alpha_{j}^{l}+\sum_{k=1}^{n}a_{jk}=1$.
Because the left side represents unit cost, $\lambda^{f}=c_{j1}$.

Using the expression for $\lambda^{f}$, the conditions in \eqref{eq:min-cost-foc-x} and \eqref{eq:min-cost-foc-l} become
\begin{align*}
l_{j}=\frac{c_{j1}\alpha_{j}^{l}}{w}\text{ and }x_{jk}=\frac{c_{j1}a_{jk}}{p_{k}}.
\end{align*}
Using these two expressions in the unit-output condition implies
\begin{align*}
1 & =l_{j}^{\alpha_{j}^{l}}\prod_{i=1}^{n}x_{ji}^{a_{ji}}\\
 & =\left(\frac{c_{j1}\alpha_{j}^{l}}{w}\right)^{\alpha_{j}^{l}}\prod_{i=1}^{n}\left(\frac{c_{j1}a_{ji}}{p_{i}}\right)^{a_{ji}}\\
 & =c_{j1}^{\alpha_{j}^{l}+\sum_{i=1}^{n}a_{ji}}\left(\frac{\alpha_{j}^{l}}{w}\right)^{\alpha_{j}^{l}}\prod_{i=1}^{n}\left(\frac{a_{ji}}{p_{i}}\right)^{a_{ji}}\\
 & =c_{j1}\left(\frac{\alpha_{j}^{l}}{w}\right)^{\alpha_{j}^{l}}\prod_{i=1}^{n}\left(\frac{a_{ji}}{p_{i}}\right)^{a_{ji}},
\end{align*}
where the last line uses the constant-returns-to-scale assumption in \eqref{eq:technology-params}.
Inverting the above expression yields an expression for $c_{j1}$:
\begin{align*}
c_{j1} = w^{\alpha_{j}^{l}}\left(\frac{1}{\alpha_{j}^{l}}\right)^{\alpha_{j}^{l}}\prod_{i=1}^{n}\left(\frac{1}{a_{ji}}\right)^{a_{ji}}\prod_{i=1}^{n}p_{i}^{a_{ji}}.
\end{align*}
The unit cost function is therefore
\begin{equation}
C\left(\bm{p},w\right)=B_{j}w^{\alpha_{j}^{l}}\prod_{i=1}^{n}p_{i}^{a_{ji}},\label{eq:unit-cost}
\end{equation}
where $\bm{p}$ is the vector of prices, $w$ is the wage rate, and
\begin{align*}
B_{j}=\left(\frac{1}{\alpha_{j}^{l}}\right)^{\alpha_{j}^{l}}\prod_{i=1}^{n}\left(\frac{1}{a_{ji}}\right)^{a_{ji}}.
\end{align*}

\subsection{Cost-minimization implies prices are determined from the supply side}

Given a representative firm's constant-returns-to-scale technology,
the minimum cost of producing $y_{j}$ units is $y_{j}C\left(\bm{p},w\right)$,
as the unit cost function scales proportionally with output \citep[][66--67]{varian_1992}.
The firm's profit-maximizing choice of output, $y_{j}$, is therefore $p_{j}=C\left(\bm{p},w\right)$.
In addition, this choice coincides with the firm earning zero profits:
if $p_{j}=C\left(\bm{p},w\right)$, then $p_{j}y_{j}=C\left(\bm{p},w\right)y_{j}=C\left(\bm{p},w,y_{j}\right)$,
so that total revenue equals total costs.
Therefore, a representative firm in sector $j$ earns zero profit.

Given that prices satisfy the zero-profit conditions across the $n$ sectors in competitive equilibrium,
the price of good $j$ must be equal to the unit cost function of that sector in \eqref{eq:unit-cost}, so
\begin{align*}
p_{j}=B_{j}w^{\alpha_{j}^{l}}\prod_{i=1}^{n}p_{i}^{a_{ji}}.
\end{align*}
Taking logs implies
\begin{align*}
\ln p_{j}=\ln B_{j}+\alpha_{j}^{l}\ln w+\sum_{i=1}^{n}a_{ji}\ln p_{i},\text{ for }j=1,\dots,n.
\end{align*}
If $w=1$ is the numeraire, then $\ln w=0$.
Under this choice,
the system of equations can be expressed as
\begin{align*}
\left[\begin{array}{c}
\ln p_{1}\\
\ln p_{2}\\
\vdots\\
\ln p_{n}
\end{array}\right]=\left[\begin{array}{c}
\ln B_{1}\\
\ln B_{2}\\
\vdots\\
\ln B_{n}
\end{array}\right]+\left[\begin{array}{cccc}
a_{11} & a_{12} & \cdots & a_{1n}\\
a_{21} & a_{22} & \cdots & a_{2n}\\
 &  & \ddots\\
a_{n1} & a_{n2} & \cdots & a_{nn}
\end{array}\right]\left[\begin{array}{c}
\ln p_{1}\\
\ln p_{2}\\
\vdots\\
\ln p_{n}
\end{array}\right],
\end{align*}
which is an $n$-equation system in $n$ prices.
Using matrix notation, the system can be expressed as
\begin{align*}
\ln\bm{p} & =\bm{b}+\bm{A}\ln\bm{p},
\end{align*}
where $\bm{b} = \left( \ln B_{1},\dots,\ln B_{n} \right)\tp$.
Prices then satisfy
\begin{equation}
  \label{eq:eqm-prices}
\ln\bm{p} =\left(\bm{I}-\bm{A}\right)^{-1}\bm{b}.
\end{equation}
This result is summarized below.

\begin{result}
  \label{result:prices}
The expression in \eqref{eq:eqm-prices} establishes prices are determined independently of the price of crude oil.
Instead, prices are determined by the supply side of the regional economy,
a standard result  in this class of models \citep{carvalho_tahbaz-salehi_2019}.  
\end{result}

\subsection{Households' choices about labor and consumption}

A representative household maximizes utility, parameterized in \eqref{eq:preferences},
subject to the budget constraint in \eqref{eq:hh-budget-constraint}.
The Lagrangian for the household is
\begin{align*}
L^{hh}=\ln v\left(l\right)+\sum_{i=1}^{n}\beta_{i}\ln c_{i}+\lambda^{h}\left[wl+\hhincome-\sum_{i=1}^{n}p_{i}c_{i}\right].
\end{align*}

The first-order condition for $c_{i}$ implies
\begin{equation}
  \label{eq:hh-foc-ci}
p_{i}c_{i} =\frac{\beta_{i}}{\lambda^{h}}.
\end{equation}
Summing across the $n$ conditions yields an expression for $\lambda^{h}$:
\begin{align*}
\sum_{i=1}^{n}p_{i}c_{i}=\sum_{i=1}^{n}\frac{\beta_{i}}{\lambda^{h}}=\frac{1}{\lambda^{h}}\sum_{i=1}^{n}\beta_{i}=\frac{1}{\lambda^{h}}.
\end{align*}
Since $\sum_{i=1}^{n}p_{i}c_{i}=wl+\hhincome$,
\begin{equation}
  \label{eq:lambda-h}
\lambda^{h} =\frac{1}{wl+\hhincome}.
\end{equation}

The household's first-order condition for labor supply implies
\begin{align*}
-\frac{lv^{\prime}\left(l\right)}{v\left(l\right)} =\lambda^{h}wl.
\end{align*}
When $v\left(l\right)=\left(1-l\right)^{\lsp}$,
using the expression for $\lambda^{h}$ in \eqref{eq:lambda-h},
the labor-supply condition amounts to
\begin{align*}
-\frac{lv^{\prime}\left(l\right)}{v\left(l\right)} & =\lambda^{h}wl=\frac{wl}{wl+\hhincome}\\
\therefore\frac{l\lsp\left(1-l\right)^{\lsp-1}}{\left(1-l\right)^{\lsp}} & =\frac{wl}{wl+\hhincome}.
\end{align*}
When the wage is the numeraire ($w=1$), the latter amounts to an expression for labor supply as a function of income generated from the crude-oil sector:
\begin{equation}
\label{eq:labor-supply}
l=\frac{1-\hhincome\lsp}{1+\lsp}.
\end{equation}
Note that labor supply, given in \eqref{eq:labor-supply}, is decreasing in income generated from the crude-oil sector, $\partial l / \partial \hhincome < 0$.
And $\partial l / \partial \lsp < 0$, as a higher $\lsp$ indicates a greater preference for leisure, $1-l$.

The first-order condition $c_{i}$ in \eqref{eq:hh-foc-ci}, $p_{i}c_{i}=\beta_{i}/\lambda^{h}$, and
the expression for $\lambda^{h}$ in \eqref{eq:lambda-h} imply
\begin{align*}
p_{i}c_{i} =\frac{\beta_{i}}{\lambda^{h}} = \beta_{i}\left(wl+\hhincome\right) = \beta_{i}\left(l+\hhincome\right),
\end{align*}
where the last equality uses the wage as the numeraire ($w=1$).
The expression for labor supply in \eqref{eq:labor-supply} implies the latter is
\begin{equation}
\label{eq:hh-summary}
p_{i}c_{i} = \frac{\beta_{i}}{1+\lsp}+\frac{\beta_{i}\hhincome \left( p_{o} \right)}{1+\lsp},
\end{equation}
where we have been explicit about how $\hhincome$ depends on the price of crude oil.

\subsection{Proof of the main result: a decomposition of changes in sectoral employment in response to a change in the price of crude oil}

In this section,
we provide a detailed derivation of result \ref{result:main-decomposition}.
This result
decomposes changes in sectoral employment in response to a change in the price of crude oil
into three effects: own-added-income, own-demand, and network-demand effects.
The decomposition works with the material-balance condition and uses optimal choices made by households and firms.

The sectoral materials-balance condition in \eqref{eq:mkt-clearing-j} is expressed in real terms.
Multiplying through by $p_{i}$ yields an expression in nominal terms for sector $i$:
\begin{align*}
p_{i}y_{i} =p_{i}c_{i}+\sum_{j=1}^{n}p_{i}x_{ji}+p_{i}z_{i}\left( p_{o} \right).
\end{align*}
Totally differentiating the nominal materials-balance condition yields
\begin{align*}
d\left(p_{i}y_{i}\right) =d\left(p_{i}c_{i}\right)+\sum_{j=1}^{n}d\left(p_{i}x_{ji}\right)+d\left(p_{i}z_{i}\right).
\end{align*}

The total differential of the nominal materials-balance condition can be developed using
a representative firm's optimal choice of input given in \eqref{eq:profit-max-choice-xji},
$p_{i}x_{ji}=a_{ji}p_{j}y_{j}$, by noting that
\begin{align*}
d\left(p_{i}x_{ji}\right)=a_{ji}d\left(p_{j}y_{j}\right).
\end{align*}
Substituting this result in the developing expression for $d\left(p_{i}y_{i}\right)$ yields
\begin{align*}
d\left(p_{i}y_{i}\right) & =d\left(p_{i}c_{i}\right)+\sum_{j=1}^{n}d\left(p_{i}x_{ji}\right)+d\left(p_{i}z_{i}\right)\\
 & =d\left(p_{i}c_{i}\right)+\sum_{j=1}^{n}a_{ji}d\left(p_{j}y_{j}\right)+d\left(p_{i}z_{i}\right)\\
\therefore\frac{d\left(p_{i}y_{i}\right)}{p_{i}y_{i}} & =\frac{d\left(p_{i}c_{i}\right)}{p_{i}y_{i}}+\sum_{j=1}^{n}a_{ji}\frac{d\left(p_{j}y_{j}\right)}{p_{i}y_{i}}+\frac{d\left(p_{i}z_{i}\right)}{p_{i}y_{i}}\\
 & =\frac{d\left(p_{i}c_{i}\right)}{p_{i}y_{i}}+\sum_{j=1}^{n}\frac{p_{i}x_{ji}}{p_{j}y_{j}}\frac{d\left(p_{j}y_{j}\right)}{p_{i}y_{i}}+\frac{d\left(p_{i}z_{i}\right)}{p_{i}y_{i}},
\end{align*}
where the last equality again uses a firm's choice on inputs given in (\ref{eq:profit-max-choice-xji}).
Therefore,
\begin{align*}
\frac{d\left(p_{i}y_{i}\right)}{p_{i}y_{i}}=\frac{d\left(p_{i}c_{i}\right)}{p_{i}y_{i}}+\sum_{j=1}^{n}\hat{a}_{ji}\frac{d\left(p_{j}y_{j}\right)}{p_{j}y_{j}}+\frac{d\left(p_{i}z_{i}\right)}{p_{i}y_{i}},
\end{align*}
where
\begin{align*}
\hat{a}_{ij}=\frac{p_{j}x_{ij}}{p_{j}y_{j}},
\end{align*}
which, as noted in the main text, differs from $a_{ij}$, the typical entries of an input--output matrix.

Turning to the choices of a representative household,
the total differential of these choices, summarized in \eqref{eq:hh-summary}, is
\begin{equation}
d\left(p_{i}c_{i}\right)=\frac{\beta_{i}d\hhincome}{1+\lsp}.\label{eq:dpici}
\end{equation}
Using this result in
the developing expression for $d\left(p_{i}y_{i}\right)/\left(p_{i}y_{i}\right)$ yields
\begin{align*}
\frac{d\left(p_{i}y_{i}\right)}{p_{i}y_{i}} = 
\frac{1}{p_{i}y_{i}}\frac{\beta_{i}d\hhincome}{1+\lsp} + \frac{p_{i}dz}{p_{i}y_{i}} + \sum_{j=1}^{n}\hat{a}_{ji}\frac{d\left(p_{j}y_{j}\right)}{p_{j}y_{j}}.
\end{align*}

Result \ref{result:prices} (prices do not change when the price of crude oil changes) implies
$d \left( p_{i}y_{i} \right)/p_{i}y_{i} = d y_{i}/y_{i}$, which is $d\ln y_{i}$.
The result is a system of equations that can be written as
\begin{align*}
\left[\begin{array}{c}
d\ln y_{1}\\
d\ln y_{2}\\
\vdots\\
d\ln y_{n}
\end{array}\right]=\left[\begin{array}{cccc}
\hat{a}_{11} & \hat{a}_{21} & \cdots & \hat{a}_{n1}\\
\hat{a}_{12} & \hat{a}_{22} & \cdots & \hat{a}_{n2}\\
\\\hat{a}_{1n} & \hat{a}_{2n} &  & \hat{a}_{nn}
\end{array}\right]\left[\begin{array}{c}
d\ln y_{1}\\
d\ln y_{2}\\
\vdots\\
d\ln y_{n}
\end{array}\right]+\left[\begin{array}{cccc}
\frac{1}{p_{1}y_{1}} & 0 &  & 0\\
0 & \frac{1}{p_{2}y_{2}} &  & 0\\
 &  & \ddots\\
0 & 0 & 0 & \frac{1}{p_{n}y_{n}}
\end{array}\right]\left[\begin{array}{c}
\frac{\beta_{1}d\hhincome}{1+\lsp}+p_{1}dz_{1}\\
\frac{\beta_{2}d\hhincome}{1+\lsp}+p_{2}dz_{2}\\
\vdots\\
\frac{\beta_{n}d\hhincome}{1+\lsp}+p_{n}dz_{n}
\end{array}\right].
\end{align*}
Using the matrix definitions
\begin{align*}
\bm{\Lambda}\equiv\left[\begin{array}{cccc}
\frac{1}{p_{1}y_{1}} & 0 &  & 0\\
0 & \frac{1}{p_{2}y_{2}} &  & 0\\
 &  & \ddots\\
0 & 0 & 0 & \frac{1}{p_{n}y_{n}}
\end{array}\right],\quad\boldsymbol{P}\equiv\left[\begin{array}{c}
\frac{\beta_{1}d\hhincome}{1+\lsp}+p_{1}dz_{1}\\
\frac{\beta_{2}d\hhincome}{1+\lsp}+p_{2}dz_{2}\\
\vdots\\
\frac{\beta_{n}d\hhincome}{1+\lsp}+p_{n}dz_{n}
\end{array}\right],
\end{align*}
the system can be solved as
\begin{equation}
  \label{eq:decomp-y}
  \begin{split}
    d\ln \bm{y} & =\hat{\bm{A}}\tp d\ln \bm{y} + \bm{\Lambda} \bm{P} \\
\therefore d\ln \bm{y} & =\left(\bm{I}-\hat{\bm{A}}\tp\right)^{-1} \bm{\Lambda} \bm{P} \\
\therefore d\ln \bm{y} & =\hat{\bm{H}}\tp\bm{\Lambda} \bm{P},
  \end{split}
\end{equation}
where $\hat{\bm{H}}$ is defined in \eqref{eq:Hhat}.

Equation \eqref{eq:decomp-y} gives the output responses to changes in the price of crude oil.
The responses of employment are proportional.
A representative firm's choice of labor, given in \eqref{eq:profit-max-choice-labor}, implies
$\ln l_{j} =\ln\alpha_{j}^{l}+\ln p_{j}+\ln y_{j}$.
Result \ref{result:prices}, which states that prices are independent of the price of crude oil, implies 
$d \ln l_{j} = d \ln y_{j}$.
This is summarized below.

\begin{result}
  \label{result:labor-proportional-output}
  Employment in industry $j$ is proportional to nominal output and
  $d \ln \bm{y} = d \ln \bm{l}$,
  which means $d \ln \bm{l}$ can be used on the left side of \eqref{eq:decomp-y}.
\end{result}

We can use result \ref{result:labor-proportional-output} to carry out the multiplication on the right side of \eqref{eq:decomp-y}
to arrive at the main decomposition in result \ref{result:main-decomposition}.

Starting from \eqref{eq:decomp-y}, we have
\begin{align*}
d\ln \bm{l} = \hat{\bm{H}}\tp\left[\begin{array}{c}
\frac{1}{p_{1}y_{1}}\frac{\beta_{1}d\hhincome}{1+\lsp}+\frac{p_{1}dz_{1}}{p_{1}y_{1}}\\
\frac{1}{p_{2}y_{2}}\frac{\beta_{2}d\hhincome}{1+\lsp}+\frac{p_{2}dz_{2}}{p_{2}y_{2}}\\
\vdots\\
\frac{1}{p_{n}y_{n}}\frac{\beta_{n}d\hhincome}{1+\lsp}+\frac{p_{n}dz_{n}}{p_{n}y_{n}}
\end{array}\right].
\end{align*}
Again carrying out multiplication on the right side yields
\begin{align*}
d\ln l & =\left[\begin{array}{cccc}
\hat{h}_{11} & \hat{h}_{21} & \cdots & \hat{h}_{n1}\\
\hat{h}_{12} & \hat{h}_{22} & \cdots & \hat{h}_{n2}\\
\vdots & \vdots & \ddots & \vdots\\
\hat{h}_{1n} & \hat{h}_{2n} & \cdots & \hat{h}_{nn}
\end{array}\right]\left[\begin{array}{c}
\frac{1}{p_{1}y_{1}}\frac{\beta_{1}d\hhincome}{1+\lsp}+\frac{p_{1}dz_{1}}{p_{1}y_{1}}\\
\frac{1}{p_{2}y_{2}}\frac{\beta_{2}d\hhincome}{1+\lsp}+\frac{p_{2}dz_{2}}{p_{2}y_{2}}\\
\vdots\\
\frac{1}{p_{n}y_{n}}\frac{\beta_{n}d\hhincome}{1+\lsp}+\frac{p_{n}dz_{n}}{p_{n}y_{n}}
\end{array}\right] \\
 & =\left[\begin{array}{c}
\sum_{i=1}^{n}\left(\frac{1}{p_{i}y_{i}}\frac{\beta_{i}d\hhincome }{1+\lsp}+\frac{p_{i}dz_{i}}{p_{i}y_{i}}\right)\hat{h}_{i1}\\
\sum_{i=1}^{n}\left(\frac{1}{p_{i}y_{i}}\frac{\beta_{i}d\hhincome }{1+\lsp}+\frac{p_{i}dz_{i}}{p_{i}y_{i}}\right)\hat{h}_{i2}\\
\vdots\\
\sum_{i=1}^{n}\left(\frac{1}{p_{i}y_{i}}\frac{\beta_{i}d\hhincome }{1+\lsp}+\frac{p_{i}dz_{i}}{p_{i}y_{i}}\right)\hat{h}_{in}
\end{array}\right].
\end{align*}
The impact of a change in the price of crude oil on employment in sector $j$ is therefore
\begin{align*}
d \ln l_{j} = \sum\limits_{i=1}^{n} \left( \frac{\beta_{i} \times d \hhincome}{p_{i}y_{i}} \times \frac{1}{1+\lsp} + \frac{p_{i}}{p_{i}y_{i}} \times dz_{i} \right),
\end{align*}
which establishes result \ref{result:main-decomposition}.

\section{Derivation of results for the $3$-sector economy}
\label{sec:app:example-economy}

In this section, we provide a derivation of the results discussed in section \ref{sec:example-economies}.

We will work with a representative firm's cost function.
The Lagrangian associated with this problem is
\begin{align*}
\mathcal{L}=wl_{i}+p_{j}x_{ij}+\psi\left(y-l_{i}^{\alpha_{i}^{l}}x_{ij}^{a_{ij}}\right),
\end{align*}
where
$\psi$ is the multiplier associated with the production constraint that $l_{i}^{\alpha_{i}^{l}}x_{ij}^{a_{ij}}\geq y$.
Taking the ratio of the two first-order conditions (to eliminate $\psi$) yields
an expression for the cost-minimizing input in terms of the choice of labor
\begin{align*}
x_{ij}=\frac{w}{p_{j}}\frac{a_{ij}}{\alpha_{i}^{l}}l_{i}.
\end{align*}
Using this condition in \eqref{eq:ex:prod} yields an expression for the conditional demand for labor,
\begin{align*}
l_{i}=y\left(\frac{p_{j}}{w}\frac{\alpha_{i}^{l}}{a_{ij}}\right)^{a_{ij}}.
\end{align*}
Using these two conditions in the cost function, $C_{i}=wl_{i}+p_{j}x_{ij}$,
and setting $y=1$ yields the unit cost function
\begin{align*}
C_{i}\left(\boldsymbol{p},w\right)=\mu_{i}w^{1-a_{ij}}p_{j}^{a_{ij}},
\end{align*}
where $\mu_{i}=a_{ij}^{-a_{ij}}\left(\alpha_{i}^{l}\right)^{-\alpha_{i}^{l}}$. 

Because the production function exhibits constant returns to scale,
costs are linear in the level of output \citep[][66--67]{varian_1992}.
In competitive equilibrium, price equals marginal cost (which, in this case, is the unit cost function), so that
\begin{align*}
p_{i}=C_{i}=\mu_{i}w^{\alpha_{i}^{l}}p_{j}^{a_{ij}}.
\end{align*}

Using the wage as the numeraire, $w=1$, the price system is
\begin{align*}
p_{i} & =\mu_{i}w^{\alpha_{i}^{l}}p_{j}^{a_{ij}}=\mu_{i}p_{j}^{a_{ij}}\\
p_{j} & =\mu_{j}w^{\alpha_{j}^{l}}p_{k}^{a_{jk}}=\mu_{j}p_{k}^{a_{jk}}\\
p_{k} & =\mu_{k}w^{\alpha_{k}^{l}}p_{i}^{a_{ki}}=\mu_{k}p_{i}^{a_{ki}}.
\end{align*}
Solving the system of equations yields
\begin{align*}
p_{i}=\gamma^{\frac{1}{1-a_{ij}a_{jk}a_{ki}}}, \quad \gamma \coloneqq \mu_{i}\mu_{j}^{a_{ij}}\mu_{k}^{a_{ij}a_{jk}},
\end{align*}
which is independent of the price of crude oil (compare result \ref{result:prices}). 

Starting from the material-balance condition $y_{i}=c_{i}+x_{ki}+z_{i}$,
multiplying by $p_{i}$ yields
\begin{align*}
p_{i}y_{i}=p_{i}c_{i}+p_{i}x_{ki}+p_{i}z_{i}.
\end{align*}
Letting a tilde above a variable denote its nominal value, the total derivative of the latter is
\begin{align*}
d\tilde{y}_{i}=d\tilde{c}_{i}+d\tilde{x}_{ki}+d\tilde{z}_{i}.
\end{align*}
To recast the term $d\tilde{x}_{ki}$ into changes in terms of output,
the representative firm's profit-maximizing choice of input, expressed in \eqref{eq:profit-max-choice-xji}, implies
\begin{align*}
p_{i}x_{ki}=a_{ki}p_{k}y_{k} \text{ or } d\tilde{x}_{ki}=a_{ki}\tilde{y}_{k}.
\end{align*}
Using this result in the developing expression associated with the nominal material-balance condition yields
\begin{align*}
d\tilde{y}_{i}=d\tilde{c}_{i}+a_{ki}\tilde{y}_{k}+d\tilde{z}_{i}.
\end{align*}
The result in \eqref{eq:dpici} applies for $d\tilde{c}_{i}$, and therefore
\begin{align*}
d\tilde{y}_{i}=\frac{d\hhincome}{3\left(1+\nu\right)}+a_{ki}\tilde{y}_{k}+d\tilde{z}_{i}.
\end{align*}

The system of equations is
\begin{align*}
d\tilde{y}_{i} & =\frac{d\hhincome}{3\left(1+\nu\right)}+a_{ki}\tilde{y}_{k}+d\tilde{z}_{i}\\
d\tilde{y}_{k} & =\frac{d\hhincome}{3\left(1+\nu\right)}+a_{jk}\tilde{y}_{j}+d\tilde{z}_{k}\\
d\tilde{y}_{j} & =\frac{d\hhincome}{3\left(1+\nu\right)}+a_{ij}\tilde{y}_{i}+d\tilde{z}_{j}.
\end{align*}
Define the term common across the three equations as
\begin{align*}
T_{\hhincome}\coloneqq\frac{d\hhincome}{3\left(1+\nu\right)}.
\end{align*}
Recursive substitution allows us to write
\begin{align*}
d\tilde{y}_{i} & =T_{\hhincome}+a_{ki}\tilde{y}_{k}+d\tilde{z}_{i}\\
 & =T_{\hhincome}+a_{ki}\left(T_{\hhincome}+a_{jk}\tilde{y}_{j}+d\tilde{z}_{k}\right)+d\tilde{z}_{i}\\
 & =T_{\hhincome}\left(1+a_{ki}\right)+a_{ki}a_{jk}\tilde{y}_{j}+a_{ki}d\tilde{z}_{k}+d\tilde{z}_{i}\\
 & =T_{\hhincome}\left(1+a_{ki}\right)+a_{ki}a_{jk}\left(T_{\hhincome}+a_{ij}\tilde{y}_{i}+d\tilde{z}_{j}\right)+a_{ki}d\tilde{z}_{k}dp_{o}+d\tilde{z}_{i}\\
 & =T_{\hhincome}\left(1+a_{ki}+a_{ki}a_{jk}\right)+a_{ki}a_{jk}a_{ij}\tilde{y}_{i}+a_{ki}a_{jk}d\tilde{z}_{j}+a_{ki}d\tilde{z}_{k}+d\tilde{z}_{i}.
\end{align*}

Therefore,
\begin{equation}
\label{eq:ex:dyi}
d\tilde{y}_{i}=\frac{1}{1-a_{ki}a_{jk}a_{ij}}\left\{ T_{\hhincome}\left(1+a_{ki}+a_{ki}a_{jk}\right)+a_{ki}a_{jk}d\tilde{z}_{j}+a_{ki}d\tilde{z}_{k}+d\tilde{z}_{i}\right\} .
\end{equation}
This expression is directly related to the response of labor.
Competitive markets and motives about profit maximization imply workers earn the value of the marginal product of labor:
$w =\alpha_{j}^{l}p_{j}y_{j} / l_{j}$, which implies $d\tilde{l}_{j}=\alpha_{j}^{l}\tilde{y}_{j}$ or $dl_{j} = \alpha_{j}^{l}\tilde{y}_{j}$,
using $w=1$.
This result together with the expression in \eqref{eq:ex:dyi} yields \eqref{eq:dl1-example}, the expression in the main text. 

\bibliographystyle{../../../bibliography/bostonfed}
\bibliography{../../../bibliography/bibliography-org-ref}

\end{document}